\documentclass[fleqn,usenatbib,useAMS]{mnras}
\usepackage{newtxtext,newtxmath}
\usepackage[T1]{fontenc}
\usepackage{ae,aecompl}
\usepackage{graphicx}
\usepackage{amsmath}
\usepackage{braket}
\usepackage{float}
\usepackage[x11names,table]{xcolor}
\usepackage{booktabs}
\usepackage{times}
\usepackage[usenames,dvipsnames]{pstricks}
\usepackage{color}
\usepackage{subfig}
\usepackage{epstopdf}

\definecolor{airforceblue}{rgb}{0.36, 0.54, 0.66}
\definecolor{royalblue(traditional)}{rgb}{0.0, 0.14, 0.4}
\definecolor{richcarmine}{rgb}{0.84, 0.0, 0.25}
\definecolor{tomato}{rgb}{1.0, 0.39, 0.28}
\definecolor{teal}{rgb}{0.0, 0.5, 0.5}
\definecolor{smalt(darkpowderblue)}{rgb}{0.0, 0.2, 0.6}
\definecolor{amber(sae/ece)}{rgb}{1.0, 0.49, 0.0}
\definecolor{applegreen}{rgb}{0.55, 0.71, 0.0}
\definecolor{brightgreen}{rgb}{0.4, 1.0, 0.0}
\definecolor{brightmaroon}{rgb}{0.76, 0.13, 0.28}
\definecolor{coral}{rgb}{1.0, 0.5, 0.31}
\definecolor{brickred}{rgb}{0.8, 0.25, 0.33}
\definecolor{brightcerulean}{rgb}{0.11, 0.67, 0.84}
\definecolor{cerulean}{rgb}{0.0, 0.48, 0.65}
\definecolor{cyan(process)}{rgb}{0.0, 0.72, 0.92}

\newcommand{\esp}{\,\,\,}

\newcommand{\lp}{\left(}
\newcommand{\rp}{\right)}

\newcommand{\be}{\begin{equation}}
\newcommand{\ee}{\end{equation}}
\newcommand{\bse}{\begin{subequations}}
\newcommand{\ese}{\end{subequations}}
\newcommand{\bary}{\begin{eqnarray}}
\newcommand{\eary}{\end{eqnarray}}
\newcommand{\beq}{\begin{equation*}}
\newcommand{\eeq}{\end{equation*}}

\newcommand{\td}{\times10}

\author[G. Morales and N. Fraija]{
G. Morales,\thanks{E-mail: gmorales@astro.unam.mx}
N. Fraija ,\thanks{E-mail: nifraija@astro.unam.mx}
\\
$^{1}$Instituto de Astronom\' ia, Universidad Nacional Aut\'onoma de M\'exico, Circuito Exterior, C.U., A. Postal 70-264, 04510, Ciudad de M\'exico, M\'exico.\\
}

\date{Accepted XXX. Received YYY; in original form ZZZ}

\pubyear{2021}
\begin{document}
\title{Neutrino propagation in winds around the central engine of sGRB}
\label{firstpage}
\pagerange{\pageref{firstpage}--\pageref{lastpage}}
\maketitle
\begin{abstract}
Since neutrinos can escape from dense regions without being deflected, they are promising candidates to study the new physics at the sources that produce them. 
{With the increasing development of more sensitive detectors in the coming years, we will infer several intrinsic properties from incident neutrinos.}
In particular, we centralise our study in those produced by thermal processes in short gamma-ray bursts (sGRBs)  and their interactions within the central engine's anisotropic medium. On the one hand, we consider baryonic winds produced with a strong magnetic contribution, and on the other hand, we treat only neutrino-driven winds. 
First, we obtain the effective neutrino potential considering both baryonic density profiles around the central engine. Then, we get the three-flavour oscillation probabilities in this medium to finally calculate the expected neutrino ratios. We find a stronger angular dependence on the expected neutrino ratios, which, incidentally, contrast from the expected theoretical ratios without considering the winds' additional contribution. The joint analysis of this observable, together with the sGRB ejected jet angle,  {might} lead to an effective mechanism to discriminate between the involved merger progenitors (black hole-neutron star or neutron star-neutron star), { acting as an additional detection channel to gravitational waves.}

\end{abstract}




\section{Introduction}

The origin of gamma-ray bursts (GRBs) remained hidden for many years since their discovery in 1967 \citep{kle73},  even when several progenitor models were developed, pretending to explain the involved physical processes \citep[see e.g.,][]{hig90,nar92}. After collecting a large sample of GRB, first shown by the BATSE experiment \citep{mee92}, and pursued the missions that followed after, it was clear that a bimodal distribution corresponding to their prompt emission duration in the $\gamma$-ray band arose. On one side, there were a set of events whose statistical distribution was concentrated over the hundreds-of-milliseconds scale. On the other hand, there were another group of events prevailing located around the tens-of-seconds timescale \citep[e.g., see Fig. 1 in ][]{ghi15,sha15}.  This bimodal behavior led to a widely accepted population classification with a chosen separation at $\sim 2$ s line \citep{kou93}, short GRB (sGRB), those whose timescale lasts less than two seconds while long GRB (lGRB) correspond to those events with a duration greater than two seconds.\\

Whereas lGRBs are associated with the collapse of a very massive star at the end of their lives (collapsar model) \citep[see][and references therein]{woo93,mac99,hjo03, woo06, hjo12}, it was not until discovering the multi-wavelength behaviour of the GW170817/GRB 170817A event that a collision of binary compact-object type was established as the responsible for creating sGRBs \citep{eichler1989nucleosynthesis,ruf01,lee04,lee05,lee2007progenitors,nak07, abott17, 2019ApJ...871..123F,2019ApJ...871..200F,2020ApJ...896...25F}. However, the study of sGRBs is far from over, leaving many unknowns to solve. In that sense, neutrinos represent a worthwhile opportunity to characterise the remaining (and sometimes hidden) GRB sources, since many of them are produced in the newly born post-merger disc. { and becomes relevant considering the fact that recent works has discussed the possible detectability of thermal neutrinos from this sources on future Megaton detectors \citep{ton16,kyu18}}. {It is worth noting that although neutrinos correspond to the least sensitive channel compared to electromagnetic and gravitational ones, these could provide additional information to constrain the progenitor's position where medium to photons is opaque. Likewise, gravitational waves detection present some disadvantages in terms of the accurate location, since they are only able to determine a large region of source location probability. So to detect point targets they need the help of other multi-wavelength telescopes to correlate the potential sources.} In the past, neutrinos from astrophysical sources have already been detected, being the multiple MeV-neutrinos from the SN1987A, the first detected particles correlated with a particular point source \citep{SN1987A}. It represented the beginning of a new era in the multi-messenger (photons and neutrinos) scenario. Even more recently, the jointly gravitational wave detection with their electromagnetic counterpart in 2017 by the Advanced LIGO Collaboration \citep[aLIGO;][]{abott17} strengthened the idea that in the coming years, a new branch in astronomy will change the way we observe our Universe.\\

Although many studies have discussed the neutrino detectability involved neutrino physics within the plasma-photon fireball during a GRB formation \citep{bah00,vol00,koe05,sah09,sah10,car20}, even including the magnetic field amplification contribution \citep{fra14,fra16}, none has considered the supplementary effect of the wind produced during the central remnant formation, in part because the density profiles were not got until recently. Therefore, in this paper, we use the multi-wavelength nature of sGRBs to explore some of these unknowns through the study of thermal neutrino properties on this medium. Particularly, we focus on neutrinos propagating within the circumburst baryon-loaded winds arising after the compact-object system took place. \\

In Section \ref{sec:grb}, we discuss the primary mechanisms of energy extraction to explain the two different approaches by which baryonic winds are driven out during the post-merger object formation. The adiabaticity condition and flip probability of stellar-like density profiles are also shown. Following that guideline, we essentially present in Section \ref{sec:neutrinos}, a brief introduction to the neutrino formalism within a particle oscillation perspective. We also indicate neutrino production's main processes and their neutrino flavour ratios. Likewise,   In Section \ref{sec:results}, we present how these winds lead to non-negligible consequences, such as deviations in the predicted oscillation probabilities, shifts in resonance energies, and most importantly, a different neutrino expected ratio, and that otherwise, it would not expect to observe. Finally, our conclusions are presented in Section \ref{sec:conclusion}. It is worth mentioning that we adopt the natural unit convention ($k_B=c=\hbar=1$).

\section{Short Gamma-ray bursts and wind formation}\label{sec:grb}

\begin{figure}
    \centering
    \includegraphics{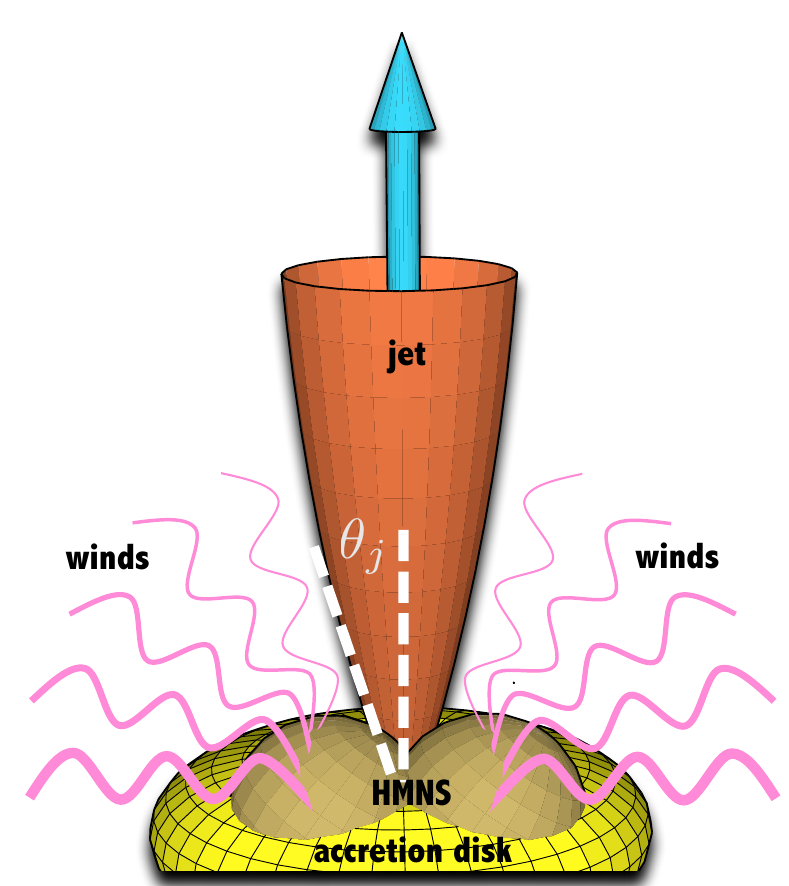}
    \caption{Schematic view  of the postmerger object (not to scale). In yellow, we display the accretion disc. It is within this hot, thick, and differentially rotating disc where the initial thermal neutrinos ($\sim$ 50 MeV) are produced. They escape through the funnel formed around polar latitudes, heating and launching baryon-loaded winds in an anisotropic distribution (pink lines). For simplicity, the rotational axis is oriented on the jet propagation direction. It is worth mentioning that latitudinal angles are treated as the half-opening jet angles. Accordingly, the <<polar angles>> refer to angles close to $\theta_j\approx0\ ^\circ$, whereas <<equatorial angles>> refer to $\theta_j\approx90\ ^\circ$.}

    \label{fig:render}
\end{figure}

The stars participating in sGRB production are the remaining compact objects from massive stars' death in a binary configuration, black hole-neutron star or neutron-neutron star (hereafter, BH-NS and NS-NS, respectively). Because of angular momentum and radiation losses in gravitational waves, these star-type objects collide with each other on a temporal scale of milliseconds \citep{rosII}, generating a coalescence with exotic physical properties. When the compact objects are two NSs, Kelvin-Helmholtz instabilities arise at the precise moment when both stars are close enough, promoting the amplification of the magnetic field by up to five orders of magnitude on the contact surface \citep{dun92,Price719,zra13,kiu14,kiu15}. The result is a hyper massive neutron star (HMNS) surrounded by an accretion disc composed from the debris of both stars that can stand stable during a relatively long time or collapse into a black hole \citep{shi06,bai08}. During this process, baryonic winds are expelled outwards in a preferential direction towards the equatorial plane, forming a cloud density that initially can be opaque (see Figure \ref{fig:render}).
Moreover, for BH-NS mergers, the BH destroys the neutron star by tidal forces generating a tail from the stellar debris. Although the result will regularly be a black hole, several simulations have proved the existence of a hot, temporary, and deferentially rotating disc preceding the black hole formation, where these winds are also produced \citep{lee2007progenitors}. In both cases, an accretion disc is formed around the progenitor with a vertical scale proportional to its radial size, temperatures of $T = 4$ MeV, and a period of rotation of $\sim$ 1 ms.\\

\subsection{Energy extraction mechanisms}

Several extraction mechanisms have been developed in recent years to explain the more feasible cooling mechanisms for the post-burst debris. In that context, some authors have considered neutrino dominated accretion flows (NDAF) as responsible for releasing a large amount of energy through neutrino emission via $\nu\bar{\nu}\to e^+e^-$ processes in low-density regions \citep{pop99,nar01,dim02}. In contrast, many others rely on MHD processes regarding the short timescales \citep{che01,koi08}. These mechanisms can be summarised into these two main groups, and we present them below. 

\subsubsection{$\nu\bar{\nu}-$ annihilation}

In this scenario, neutrinos with energies $\sim $ 50 MeV are produced by pairs annihilation (cf. Section \ref{subsec:production}) within the accretion disc (or around it). Initially, the disc is optically thick (reaching $ \tau \sim10^4 $), and neutrinos look for a low-density region usually located in the funnel formed along the rotation axis; once there, the neutrino density gradient increases. Consequently, the annihilation rate grows to produce a fireball composed essentially of $(\gamma, e^\pm)$. The fireball expands relativistically due to the low amount of baryonic material it has. In turn, the outflow of these neutrinos constitutes an effective mechanism for cooling the system, which drags and heats the surrounding baryonic medium into the so-called neutrino-driven winds (hereafter, NDW), which in principle is responsible for collimating the jet \citep{ros02_jets}. It is worth mentioning that part of the created energy by neutrinos is re-deposited into the disc in a colder flow ($T = 1$ MeV), continuing to feed the winds around the progenitor.  Again, from the interactions with the baryonic material and the fireball's electrons, lower energy neutrinos are created.  In fact, \citep{rosII} found that neutrinos have an average energies of the order of ($E_\nu \simeq 8,15,20$-$25$) MeV for electronic, muonic and tauonic neutrinos, respectively. Unfortunately, the exact calculations of the wind density profiles are difficult to calculate and only recently were performed \citep[e.g., see][]{des08,siegel2014magnetically,perego2014neutrino}.

\subsubsection{MHD processes}
The disadvantage of the previous mechanism is that it results in an extremely inefficient process because it requires a large neutrino luminosity or very short timescales \citep{ruf97}; thus, other energy extraction mechanisms through MHD processes have also been proposed. That can be done either from the accretion disc during the HMNS stable phase or by rotational energy during the postmerger object collapse to a BH using the so-called Blandford-Znajek mechanism \citep{bla77,daigne02}. In the first case,  a toroidal magnetic field in the disc plane is formed, whose field lines are forced to reconnect quickly and then twisted due to the differential rotation of the disc and the magnetic field amplification. This process heats the surrounding medium, ejecting a wind outflow \citep{rosIII}. In the second case, the energy extracted by the Blandford-Znajek mechanism is injected into the newly created winds from the disc debris through the Poynting flux before being converted into gamma rays. These winds tend to follow the magnetic field lines \citep{bla82}. In all the conditions mentioned above, a very intense magnetic field ($B>10^{15}$ G) is needed. Therefore, we will refer to magnetically-driven winds (hereafter, MDW) as those produced during a BNS merger because this is the only known configuration where the magnetic field increases up to this intensity \citep{Price719,gia09,kiu14,kiu15}.

\subsection{Adiabaticity condition and flip probability}

In a medium with a variable density, the neutrino properties are modified and governed primarily by an adiabaticity condition. The concept of adiabaticity is related to how quickly the system adapts to changing external conditions. In the ideal case, we demand a smooth change in the external variable agent, in this case, the baryonic density. If the density varies smoothly as a function of distance, the neutrino states' changes can be negligible. Therefore, the flavour admixtures are preserved and can take representative values of the density. Besides, the dynamics of the transitions are determined by the adiabaticity parameter \citep{par86,dig00}

\begin{equation}
    \label{eq:adiabatic_parameter}
    \gamma_{\rm ad}\equiv \dfrac{\Delta m^2}{2 E_\nu} \dfrac{\sin^22\theta}{\cos 2\theta}\dfrac{1}{\frac{1}{n_e}\frac{dn_e}{dr}}\ ,
\end{equation}

and  the probability that a defined neutrino in a mass eigenstate jumps to another, the so-called flip probability is defined as
\begin{equation}
    \label{eq:flip_proba}
     P_f =  e^{-\frac{\pi \gamma_{\rm ad}}{2}}\ ,
\end{equation}

which is given by the linear approximation of the Landau-Zener formula. In this context, the adiabaticity condition is satisfied when $ \gamma \gg 1 $, which translates into a low flip probability. Furthermore, this parameter can be expressed as a function of a power-law density profile $ \rho = Ar^{- n} $ as follows
\begin{equation}
    \label{eq:gamma}
    \gamma=\dfrac{1}{2n}\lp \dfrac{\Delta m^2}{E_\nu}^{1-\frac{1}{n}}\rp \dfrac{\sin^22\theta}{(\cos 2\theta)^{1+\frac{1}{n}}} \lp\dfrac{2\sqrt{2}G_F Y_e}{m_N}A  \rp^{\frac{1}{n}}\ ,
\end{equation}

where the resonance condition has been used to indicate the distance contribution through the oscillation parameters, and the number density of electrons was expressed as $n_e=Y_e\rho(r)/m_N$.


\section{Neutrinos}\label{sec:neutrinos}
In particle physics, transitions and survival probabilities are obtained from the temporal evolution of the Schr\"odinger equation associated with each particle state. The calculus of these probabilities is crucial for understanding particle oscillations in different contexts. This theory has been extensively developed since its discovery; consequently, we will only outline the most important theoretical aspects to consider in this work.

\subsection{Neutrino formalism}
In the case of neutrinos, these are created by weak interactions through charged-current processes with charged leptons. A defined neutrino of flavour $\alpha$ and momentum $\vec{p}$ is described in a flavour state as $\ket {\nu_\alpha}$ with $( \alpha = e, \mu, \tau )$. Nevertheless, as soon as they propagate through space, they can be found in a juxtaposition of the so-called <<mass eigenstates>> $\ket{\nu_a}$ with $(a=1,2,3)$ \footnote{It is important to emphasize that in literature, the notation of Greek letters refer to flavour eigenstates. In contrast, Latin letters are used for mass eigenstates}, which evolve, so that a neutrino created with a  flavour $\alpha$ can be detected with a different flavour $\beta$ \citep{pon68,bar80}, mathematically this can be expressed as:

\begin{equation}\label{eq:mass_flavour_relation}
    \ket{\nu_\alpha}=\sum_{a=1}^3 U_{\alpha a}^* \ket{\nu_a}\, ,
\end{equation}
i.e., flavour eigenstates  are expressed in terms of mass eigenstates  and vice versa, through the unitary matrix $U^*$ and $U$, respectively. Assuming a flat wave approximation, the temporal evolution of neutrinos is governed by the Schr\"odinger equation $ i\ket{\dot{\nu_\alpha (t)}}=\mathcal{H}_f\ket{\nu_\alpha(t)}$, whose solution is given by 
  \begin{equation}\label{eq:Sch_solution}
      \ket{\nu_\alpha(t)}=e^{-i\mathcal{H}_f t}\ket{\nu_\alpha(t)}=\mathcal{U}_f(t)\ket{\nu_\alpha(t)}\ ,
  \end{equation}
 with $\mathcal{H}_f$ the hamiltonian  and $\mathcal{U}_f$ the temporal evolution operator in a flavour basis. It is more feasible to work on a mass eigenstate basis where  $\mathcal{H}_f$ is diagonal rather than a flavour basis where it is not. Therefore, we use a unitary matrix to perform the basis transformation, such as,

\begin{equation}
    \label{eq:hamiltonian_transformation}
    \mathcal{H}_m=U^{-1}\mathcal{H}_f U \ .
\end{equation}
In a three-flavour mixing scenario in a vacuum, this unitary matrix is known as the PMNS matrix and is parametrized in terms of the mixing angles  as \citep{mak62,gro74}

  \be
U = \begin{pmatrix}

c_{13}c_{12}                    & s_{12}c_{13}                    & s_{13}\cr
-s_{12}c_{23}-s_{23}s_{13}c_{12} & c_{23}c_{12}-s_{23}s_{13}s_{12}   & s_{23}c_{13}\cr
s_{23}s_{12}-s_{13}c_{23}c_{12}  &-s_{23}c_{12}-s_{13}s_{12}c_{23}   &  c_{23}c_{13}\cr
\label{matrixmezcla}
\end{pmatrix}\,,
\ee
with $c_x\equiv\cos\theta_x$ and $s_x\equiv \sin\theta_x$, being $x=1,2,3$ the vacuum mixing angles.\\

Finally, the transition probability amplitude is given by
\begin{align}
    \label{eq:amplitude1}
    A_{\alpha\beta}(t)\equiv&\braket{\nu_\beta|\mathcal{U}_f(t)|\nu_\alpha }\nonumber\\ =& \sum_{a=1}^n\sum_{b=1}^n U_{\alpha a}^* U_{\beta b} e^{-i(\mathcal{H}_{m,b}-\mathcal{H}_{m,a})\ t}\braket{\nu_b|\nu_a}\ .
\end{align}
 In the vacuum, the hamiltonian only depends on the 4-momentum given by $\mathbf{P_a}\equiv (E_a,\vec{p}_a)=(E_a,p_a,0,0)$, where $E_a=\sqrt{m_a^2+\vec{p}_a^2}=p_a\sqrt{1+(m_a/p_a)^2}\approx p_a+\frac{m_a^2}{2p_a}$ are the neutrino energies for the mass eigenstates $\ket{\nu_a}$, $m_a$ are the neutrino masses, and $\vec{p}$ represents the 3-momentum vector $(p_a,0,0)$. Then, we can rewrite Equation (\ref{eq:amplitude1}) as
 \begin{align}
     \label{eq:amplitude2}
      A_{\alpha\beta}(t)\equiv&\sum_{a=1}^n\sum_{b=1}^n U_{\alpha a}^* U_{\beta b} e^{-i(\mathbf{P}_b-\mathbf{P}_a)\ L}\delta_{ba}\nonumber\\
      =& \sum_{a=1}^n U_{\alpha a}^* U_{\beta a} e^{-i(p_a-E_a)\ L}\\
      =& \sum_{a=1}^n U_{\alpha a}^* U_{\beta a} e^{-i(p_a-p_a-(m_a^2/2p_a))\ L}\nonumber\\
      =& \sum_{a=1}^n U_{\alpha a}^* U_{\beta a} e^{i\frac{m_a^2L}{2E_a}}\ ,\nonumber
 \end{align}
where we have used the fact that neutrinos are relativistic and consequently $t\approx L$, being $L$ the neutrino path length. Moreover, in a relativistic limit  $|\vec{p}_a\gg m_a|$, so the approximation $E_a\approx \vec{p}_a$ is still valid.\\

Finally, the transition probability is   $P_{\alpha\beta}(t)\equiv|A_{\alpha\beta}(t)|^2$,  recovering the well-known formula for neutrino oscillations in vacuum
\begin{equation}
    \label{eq:proba_vacuum}
    P_{\alpha\beta}=\delta_{\alpha\beta}-4\underset{a<b}{\sum^n_{a=1}\sum^n_{b=1}}\ \mathbb{R}{\rm e}\ J_{\alpha\beta}^{ab} \sin^2\Phi_{ab}  +2 \underset{a<b}{\sum^n_{a=1}\sum^n_{b=1}}\ \mathbb{I}{\rm m}\ J_{\alpha\beta}^{ab}  \sin(2\Phi_{ab})\ ,
\end{equation} 
with $J_{\alpha\beta}^{ab}\equiv U_{\alpha a}^* U_{\beta a} U_{\alpha b} U_{\beta b}^*$ and $\Phi_{ab}\equiv \frac{\Delta m_{ab}^2L}{4E}$. It is worth noting that for antineutrinos,  the complex term only changes sign.\\ 

From the last, the oscillation length is defined as:
\begin{equation}
    \label{eq:Losc}
    L_{ab}^{\rm osc,v} \equiv \frac{4\pi E}{\Delta m_{ab}^2}\ ,
\end{equation}
physically, the Equation (\ref{eq:Losc}) defines the minimum length from which the oscillations take place. In the MeV-range, this length turns out to be  between $1.3\times10^{6}$ and $7.6\td^7$ cm \citep{fra16}, where the difference lies  on the mixing parameters taking into account. Additionally, in a three-flavour scenario, there are two corresponding resonance energies associated with the electron density in the medium. These are defined as \citep{raz10}
\begin{equation}
    \label{eq:energy_res}
    E_{\rm res}^{L}\approx\frac{\Delta m_{21}^2}{2V_{\rm eff}}\cos2\theta_{12}\,\esp\esp \esp\esp \esp\esp     E_{\rm res}^{H}\approx\frac{\Delta m_{31}^2}{2V_{\rm eff}}\cos2\theta_{13}\,
\end{equation}


These energies define regions with different transition characteristics. For $E_\nu< E_{\rm res}^{L}$ neutrinos oscillate in vacuum lookalike medium, while for $E_\nu> E_{\rm res}^{L}$ the matter effects become dominant. This is because for neutrinos propagating in a non-vacuum medium, the additional matter contributions modify the Hamiltonian. In order to consider the matter effect, we can express the Hamiltonian in the mass basis for a three-flavour mixing scenario as \citep[e.g., see][]{2014MNRAS.442..239}

\begin{equation}
    \mathcal{H}_{m,\rm mat}=\mathcal{H}_{m,\rm vac}+U^{-1}\mathcal{V}_f U\,,
\end{equation}
with $\mathcal{H}_{m,\rm vac}$ and $\mathcal{V}_f$ in a three-flavour scenario  given by 
\begin{equation}\label{eq:H_m}
    \mathcal{H}_{m,\rm vac}=\begin{pmatrix}
E_1 & 0 &0 \\ 
 0& E_2 &0 \\ 
0 & 0 & E_3 
\end{pmatrix}\,,\esp\esp \mathcal{V}_f=\begin{pmatrix}
V_{\rm eff} & 0 &0 \\ 
 0& 0 &0 \\ 
0 & 0 & 0 
\end{pmatrix}\,,
\end{equation}
where $V_{\rm eff}$ is the charged-current potential
\begin{equation}\label{eq:Vcc}
    V_{\rm eff}=\sqrt{2}\ G_F\ N_e(r)\simeq \frac{\sqrt{2}\ G_F\ Y_e \rho(r)}{m_N}\,,
\end{equation}
with $G_F$ the Fermi constant, $N_e$ the number density of electrons, $Y_e$ the average electron fraction, $m_N=(m_p+m_n)/2$ the nucleon mass, $m_p$ the proton mass, $m_n$ the neutron mass, and $\rho(r)$ the mass density.\\

The Hamiltonian in Equation (\ref{eq:H_m}) is not diagonal either in the flavour or mass basis, so the transition probabilities can only be obtained from the evolution operator $\mathcal{U}_f=e^{-i\mathcal{H}_{f,\rm mat}\ L}$ with $\mathcal{H}_{f,\rm mat}=U\ \mathcal{H}_{m,\rm mat}\ U^{-1}$, such as, 
\begin{equation}
    \label{eq:Proba_m}
    P_{\alpha\beta}=A_{\alpha\beta}^2=|\braket{\nu_\beta|\mathcal{U}_f|\nu_\alpha}|^2\ .
\end{equation}

 In this scenario, calculations of $\mathcal{U}_f$ become quickly complicated, and even when approximate and analytical expressions within different contexts have been obtained previously for particular cases \citep[see, e.g.,][]{bar80,hax87,kuo87,dol96,bal96,ohl00,akh04,glz03}, it is usual to obtain these transition probabilities numerically.

\subsection{Neutrino ratio parametrization}
Once the probability expressions are got, we can determine the fraction of neutrino flavours that leave the source through the following relation
\begin{equation}
    \label{eq:F_nu}
     F = P_{\alpha\beta} F_0  \ ,
\end{equation}

with $P_{\alpha\beta}$ the matrix of transition probabilities Equation (\ref{eq:Proba_m}). $F_0\equiv(\nu_e,\nu_\mu,\nu_\tau)^T_{\rm created}$ and $F\equiv (\nu_e,\nu_\mu,\nu_\tau)^T$  are the neutrino flux vectors before and after oscillations take place, respectively. Equation (\ref{eq:F_nu}) are usually parameterized to obtain the expected fraction of neutrinos. In this work we will consider w.l.o.g. the parameterization made by \citep{pal15} where  neutrino fractions are represented as

\begin{equation}
    \label{eq:xi_n}
    \xi_n=F_n/\sum_n F_n\ .
 \end{equation}
Whether the initial neutrino fractions are known $(\xi_e,\xi_\mu,\xi_\tau)_{\rm created}\equiv(f,g,h)$,  then  neutrino flavour ratios after propagation can be parameterized as:

\begin{align}
    \label{eq:xi}
    \xi_e=&\frac{1}{3}+(2-3g-3h)P_0+(g-h)P_1\ ,\nonumber\\
     \xi_\mu=&\frac{1}{3}+\frac{1}{2}(-2+3g+3h)P_0\\
     &+(1-2g-h)P_1+(g-h)P_2\ ,\nonumber\\
     \xi_\tau=&\frac{1}{3}+\frac{1}{2}(-2+3g+3h)P_0\nonumber\\
     &+(-1+g+2h)P_1-(g-h)P_2\ ,\nonumber\\
\end{align}

with $P_0$ $P_1$ and $P_2$  expressed in term of each independent transition probability as

\begin{align}
    \label{eq:P_i_Palladino}
    P_0=&\dfrac{P_{ee}-\frac{1}{3}}{2}\ ,\nonumber\\
    P_1=&\dfrac{P_{e\mu}-P_{e\tau}}{2}\ ,\\
    P_2=&\dfrac{P_{\mu\mu}+P_{\tau\tau}-2P_{\mu\tau}   }{4}\ .\nonumber
\end{align}

\subsection{Neutrino production mechanisms}\label{subsec:production}

Due to the high temperatures peaked during the initial stage, several neutrino emissions within the plasma fireball occur. Under these conditions the neutrino mean free path is of the order of $\lambda\sim0.75$ m$(5\td^{14}$ g cm$^{-3}/\rho) \ (10$ MeV$/T)^2$ \citep{rosII}, being affected principally by the interaction with nucleons. The leading neutrino processes are  \citep{dic72,lat76}:

\begin{itemize}
\item pairs annihilation ($e^++e^-\to\nu_x+\bar{\nu}_x$),
\item plasmon decay $(\gamma\to\nu_x+\bar{\nu}_x)$,
\item photo-neutrino emission $(\gamma+e^-\to e^-+\nu_x+\bar{\nu}_x)$,
\item positron capture $(n+e^+\to p+\bar{\nu}_e)$,
\item electron capture $(p+e^-\to n+\nu_e)$,
\end{itemize}
 with ($x=e,\mu,\tau$). Since the last interaction is the only one that produces neutrinos with a defined flavour $e$, then we will consider that the initial neutrino rate in this work is $(4\nu_e,3\nu_\mu,3\nu_\tau)$.

\subsection{Neutrino global fits}
We summarize in Table \ref{table1} the most recent global fits for three-flavour neutrino admixtures in a normal-ordering (NO) and inverted-ordering (IO) configuration \citep{est19}.

\begin{table}
\caption{Shown are the best-fit parameters for a three-flavour mixing scenario considering  a NO scheme  ($\Delta m_{31}^2>0$; left column)  and an IO scheme ($\Delta m_{31}^2<0$; right column). These parameters were obtained within  $\pm1\sigma$ range from  global data analysis performed by \citep{est19}.  }
\centering \rowcolors{1}{}{applegreen!18}
\begin{tabular}{ccc}\hline\hline\\
 \rowcolor{richcarmine!40}\textbf{Parameter}                             & \textbf{Best-fit  $\pm1\sigma$ (NO)} & \textbf{Best-fit  $\pm1\sigma$ (IO)}   \\
  \hline\hline  \addlinespace[1mm]\\
$\text{sin}^2\theta_{12}$                      & ${0.310_{-0.012}^{+0.013}}$ &  ${0.310_{-0.012}^{+0.013}}$  \\  \addlinespace[1.2mm]
$\theta_{12}/\ ^\circ$                          & $33.82_{-0.76}^{+0.78}$    & $33.82_{-0.75}^{+0.78}$  \\  \addlinespace[1.2mm]
$\text{sin}^2\theta_{23}$                      & $0.582_{-0.019}^{+0.015}$   & $0.582_{-0.018}^{+0.015}$ \\  \addlinespace[1.2mm]
$\theta_{23}/\ ^\circ$                          & $49.7_{-1.1}^{+0.9}$       & $49.7_{-1.0}^{+0.9}$      \\  \addlinespace[1.2mm]
$\text{sin}^2\theta_{13}$                      & $0.02240_{-0.00066}^{+0.00065}$ & $0.02263_{-0.00066}^{+0.00065}$\\ \addlinespace[1.2mm]
$\theta_{13}/\ ^\circ$                          & $8.61_{-0.13}^{+0.12}$    &  $8.65_{-0.13}^{+0.12}$  \\ \addlinespace[1.2mm]
$\dfrac{\Delta m_{21}^2}{10^{-5}\text{ eV}^2}$ & $7.39_{-0.20}^{+0.21}$     & $7.39_{-0.20}^{+0.21}$    \\ \addlinespace[1.2mm]
$\dfrac{\Delta m_{31}^2}{10^{-3}\text{ eV}^2}$ & $2.525_{-0.031}^{+0.033}$  &  $-2.512_{-0.031}^{+0.034}$  \\ \addlinespace[1.2mm] \hline\hline
\end{tabular}
\label{table1}
\end{table}

\section{Results
}\label{sec:results}


{From Equation 15, we recognise that the potential depends  on the electron number density for each configuration.  Accordingly, we infer that the electron fraction plays an important role within the postmerger remnant ejection and that it is far from being uniform in these scenarios, since the development of the winds depends especially on the progressive evolution of this value. Because of this, many works have discussed this issue, for example, \cite{perego2014neutrino} simulated the spatial distribution of this electron fraction for winds ejected in the close vicinity of the remnant (without considering magnetic contributions).   In these simulations, the authors concluded that indeed, both the matter in the nearest parts of the disk and that surrounding the HMNS before being ejected, have different values. In particular, they mention that for high latitudes ($50^\circ-90^\circ$), the value of $Y_e$ is expected to be close to the value of the equilibrium electron fraction $Y_{\rm e,\ EQ}$, while for low latitudes ($30^\circ-50^\circ$) its value typically ranges between $0.2$ and $0.4$, although an increase to a value close to $Y_{\rm e,\ EQ}$ expected but on a longer time scale. Undoubtedly, this contributes to a clarification of how the ejected baryonic mass is distributed and how it evolves over time. Nevertheless, we are aware that a small variation in this value will have a negligible effect on the associated potential and consequently on the oscillation parameters, even for extreme angular values. For this reason, in this work we do not consider further any temporal or spatial variations of the value of $Y_e$ within the winds.} \\

In order to identify the  angular variation of the postmerger winds density, we rely on the work made by \citep{mur17}. They obtained the baryonic density profiles from two global simulations considering two different mechanisms of wind production \citep{siegel2014magnetically,perego2014neutrino}. In the first one, HD simulations were performed to characterise the winds that are expelled by neutrinos. In contrast, in the second work, they study the winds surrounding the progenitor in a strongly magnetised environment  ($B=10^{14}-10^{16}$ G) as a result of a BNS collision.\\

\begin{figure}
    \centering
    \includegraphics[width=0.45\textwidth]{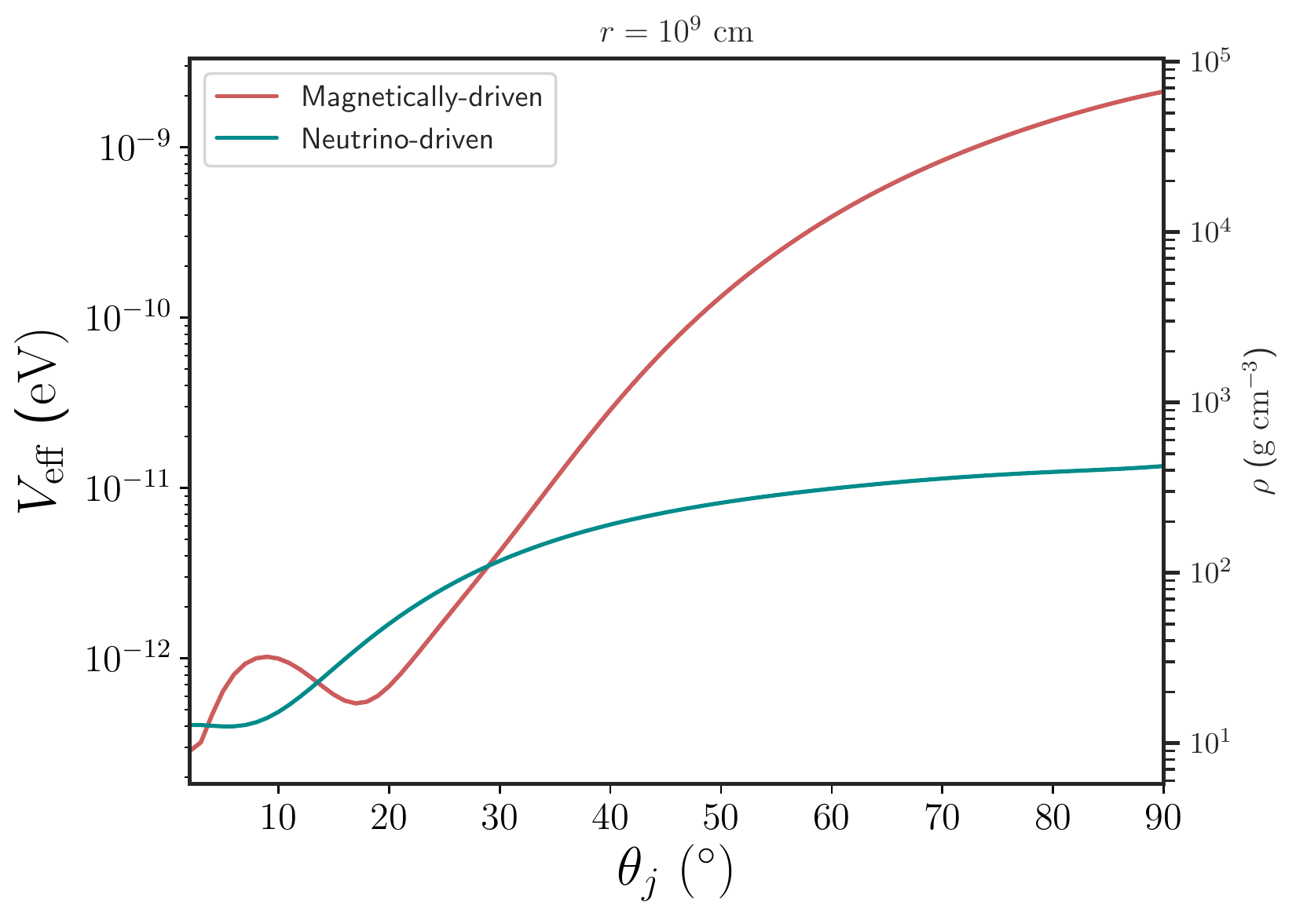}
    \caption{Latitudinal density profiles calculated at $ r = 10^9 $ cm of \citep{mur17}. Additionally, we display the related neutrino effective potential for each scenario. One notes that the greater contribution to potential takes place in the equatorial region.}
    \label{fig:Vcc}
\end{figure}

Considering these density profiles and using   $Y_{\rm e,\ EQ} = 0.42$ \citep{qia96}  in Equation (\ref{eq:Vcc}), we show in Figure \ref{fig:Vcc}, the density profiles and the effective potential at radius $r=10^9 $ cm for both wind conduction channels. In this Figure, we obtained the potential within the winds during both MDW and NDW launch. We found that in the magnetic case, the potential increases up to three orders of magnitude reaching a maximum value of $V_{\rm eff} = 2.125\times10^{-9}$ eV at  $\theta = 90\ ^\circ$ in comparison with  $V_{\rm eff} = 3.213\times10^{-12}$ eV at $\theta = 0\ ^\circ$. On the other hand, when we examine the NDW case, we realise that the potential does not have a substantial increment, being only $V_{\rm eff} = 1.347\times10^{-13}$ eV at $\theta = 0\ ^\circ$ and $V_{\rm eff} = 4.060\times10^{-13}$ eV at $\theta = 90\ ^\circ$ with an absolute rise of only one order of magnitude. Remarkably, the potential behaviour in both cases remain quite similar within the polar vicinity, lying on the range $(3.2\times10^{-13}\ \rm{eV}<V_{\rm eff}<3.5\times 10^{-12}\ \rm{eV})$, but it has a  meaningful increment from $\theta_j> 29\ ^\circ$. In other words, we are not able to distinguish between the two types of progenitors for an on-axis event, mainly because the neutrino oscillation properties remain similar in both cases. Conversely, for an off-axis event seeing `edgewise', the potential will have a significant contribution for greater latitudinal angles due to the magnetic amplification, and accordingly, the neutrino oscillation probabilities, as well as the flavour rates, will allow us to identify when a binary neutron star collision forms the resulting sGRB.\\

\begin{figure}
    \centering
    \includegraphics[width=0.45\textwidth]{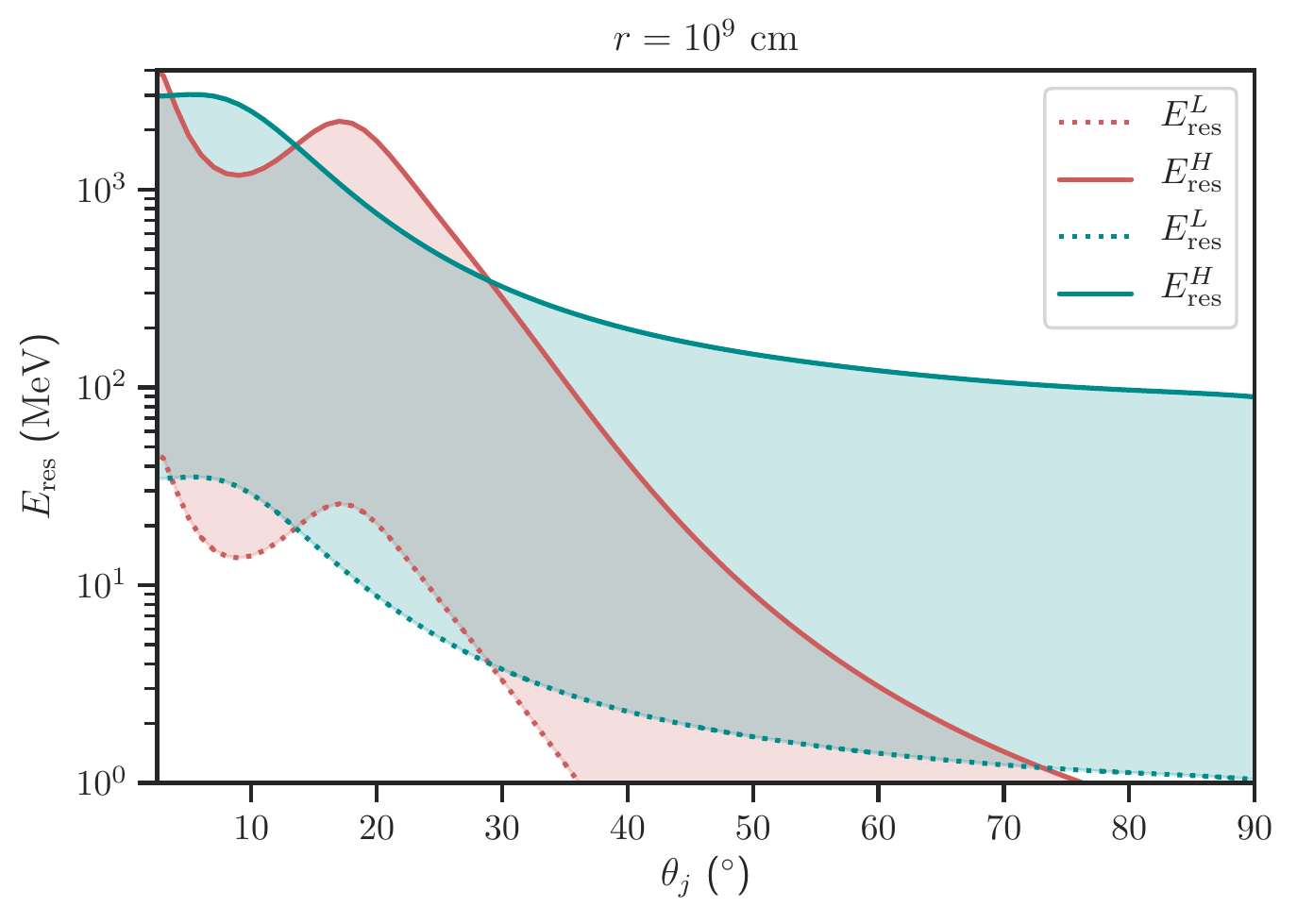}
    \caption{Resonance energies for both, MDW (red) and NDW  (blue) winds. The dotted lines describe the lower limits for the resonance energies while the continuous lines are the upper limits for our considered cases. The filled area represent the region $E_{\rm res}^L<E_\nu<E_{\rm res}^H$ where neutrinos oscillate between the two resonances and therefore the matter effects prevail.}
    \label{fig:Res_Energy}
\end{figure}

Likewise, with the densities presented above, we plot in Figure \ref{fig:Res_Energy}, the resonance energies using Equation (\ref{eq:energy_res}). We find those regions where neutrino oscillations become significant. In particular, we stand out: i) the region below $E_{\rm res}^L$ which be in agreement with the vacuum oscillation region where the admixtures between both $1\leftrightarrow2$ and $2\leftrightarrow3$ mass eigenstates dominates,  ii) the filled region $E_{\rm res}^L<E_\nu<E_{\rm res}^H$, where the matter effects dominate. It is worth mentioning that for $E_\nu>E_{\rm res}^H$, the medium effect also remains  worthwhile as long as the energy does not exceed the breakdown value of adiabaticity energy.
Following this guideline, we recognise that the resonance energies in the MDW case encompass a region between $43.697\ \rm  MeV<E\nu<3.752\  \rm  GeV$ at $\theta_j=0\ ^\circ$ and decreases until $6.606\  \rm  keV<E\nu<0.567\  \rm  MeV$ at $\theta_j=90\ ^\circ$, while for the NDW case, the resonance interval is entirely included between $34.580\  \rm MeV<E\nu<2.970 \ \rm GeV$  at $\theta_j=0\ ^\circ$ and $1.042\  \rm MeV<E\nu< 89.529\  \rm MeV$ at $\theta_j=90\ ^\circ$, and  iii) for thermal neutrinos produced within a magnetised environment, our focus is for $\theta_j>36.04\ ^\circ$ where the lower resonance energy is greater that $E_\nu=1.022$ MeV (twice the electron mass value at rest frame), being this region where  oscillations differ from the case in the vacuum. Thermal neutrinos with energies as low as the minimum energy of creation in the complementary case, behave like in the vacuum for all propagation angles but show a different performance as their energy rises. For instance, at $E_\nu=10$ MeV, resonances arises from $\theta_j>18.87\ ^\circ$, as well as, from $\theta_j>26.01\ ^\circ$ for neutrinos with $E_\nu=5$ MeV.\\

\begin{figure}
    \centering
    \includegraphics[width=0.45\textwidth]{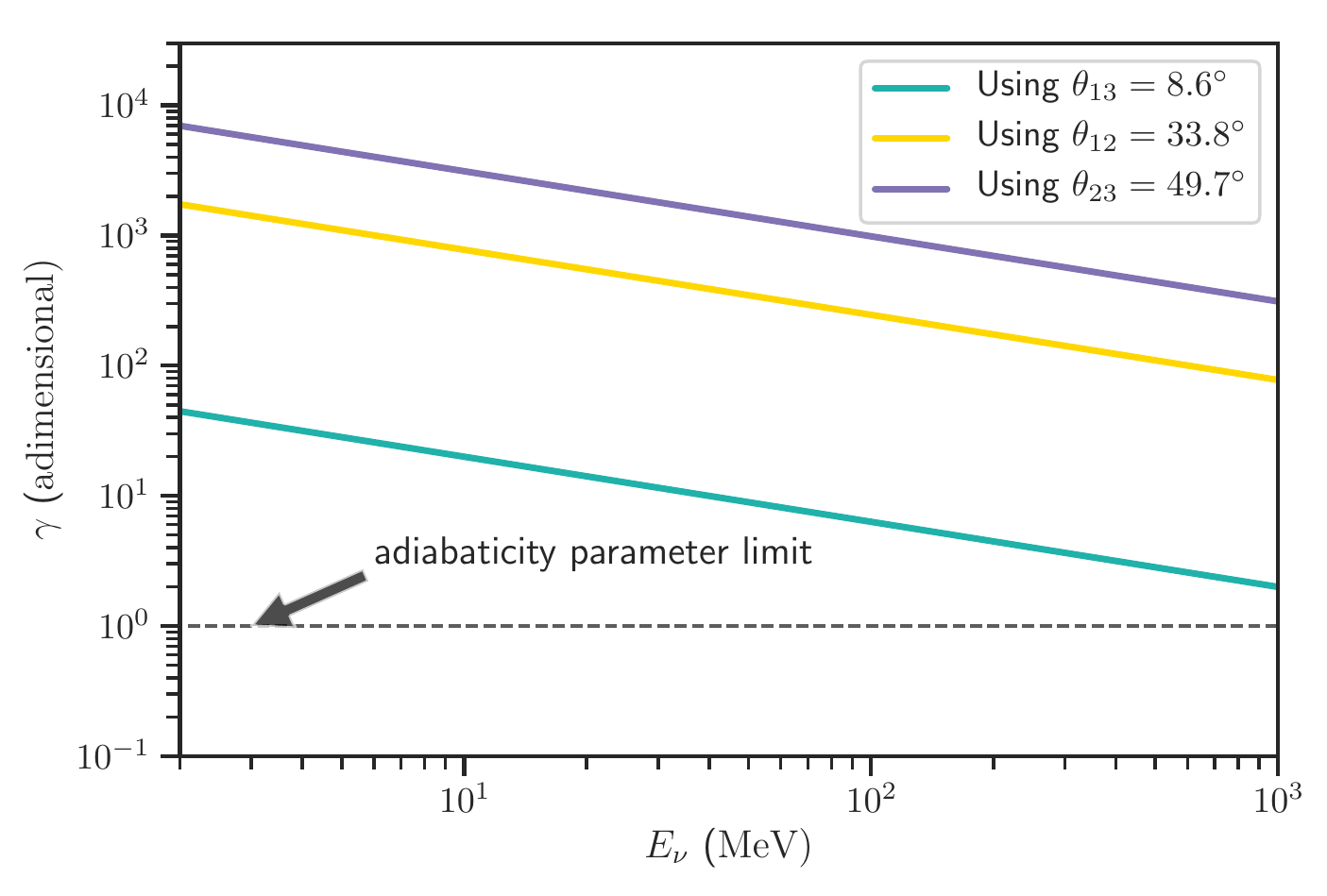}
    \caption{Adiabaticity parameter for a medium with density profile $\rho\propto  r^{-2}$ regarding the three possible admixture angles in a three-flavour scenario \citep{est19}. We observe that neutrinos with $E <1$ GeV propagate adiabatically within a circumburst stellar wind.}
    \label{fig:gamma}
\end{figure}

In order to know the adiabaticity breakdown energy, we first  acknowledge that for circumburst stellar winds, the density profile takes the form \citep{mur14}
\begin{equation}
     \label{eq:rho_r}
  \rho_w(r)=\frac{\dot{M}}{4\pi c \beta r^2}= Ar^{-2}\ ,
\end{equation}
where $A\equiv \dot{M}/(
4\pi c \beta_w)$ is the corresponding proportional constant, with $\dot{M}\simeq 10^{-3}\ M_\odot \rm s^{-1}$ and $\beta_w\simeq 0.3$
the typical values of mass loss rate and wind ejection velocity. To determine if a density profile of these characteristics is smoothly enough to be considered adiabatic, the  parameter $\gamma$ must be calculated. Therefore, using Equation (\ref{eq:gamma}), we show this parameter in Figure \ref{fig:gamma}, regarding the three possible mixing angles from Table \ref{table1} as a function of neutrino energy.\\

\begin{figure}
    \centering
    \includegraphics[width=0.45\textwidth]{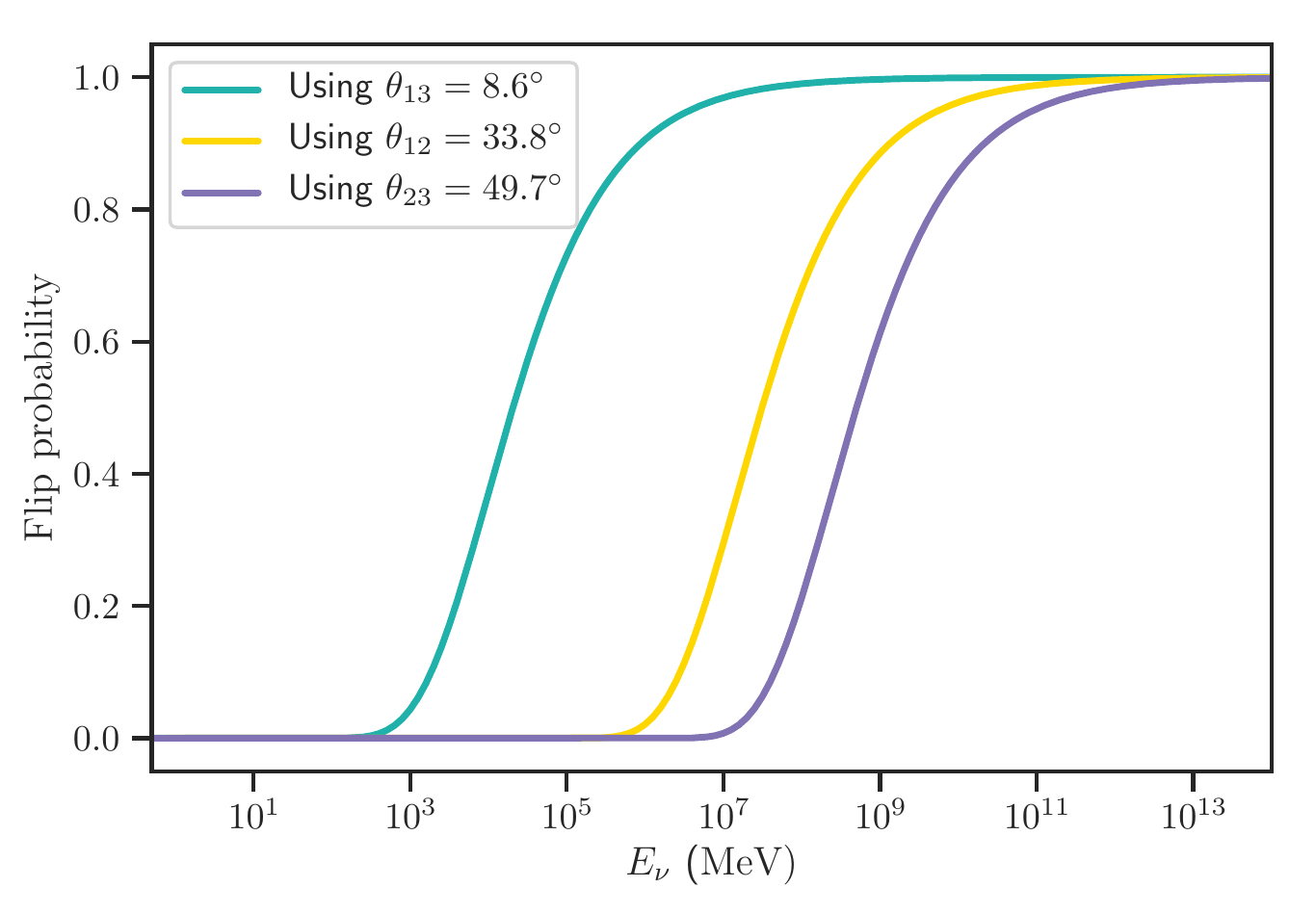}
    \caption{Flip probability obtained from the adiabaticity parameter.}
    \label{fig:flip}
\end{figure}

We find that the adiabaticity condition for all mixing angles $\gamma \gg 1$ is fulfilled for $E_\nu<1$ GeV. Furthermore, using Equation (\ref{eq:flip_proba}), we show  in Figure \ref{fig:flip} the flip probability as a function of the neutrino energy from the adiabaticity parameters previously obtained. In this Figure, we highlight three well-defined transition regions for the density profile in Equation (\ref{eq:rho_r}). i) Where the flip probability is zero; in this region is where all adiabatic conversions for neutrinos with $E_\nu \lesssim 10^3$ MeV take place, guaranteeing that for this profile, jumps between mass eigenstates are forbidden, ii) the region which $ P_f $ fluctuates between 0 and 1;  in this case, the probabilities can be approximated with the average variation $ \braket { P_f} $, and iii) where $ P_f = 1 $ implying that the adiabaticity condition is strongly violated. Since we are studying  thermal neutrinos within the MeV-range, then conversions will be completely adiabatic.\\

\begin{figure}
    \centering
    \includegraphics[width=0.45\textwidth]{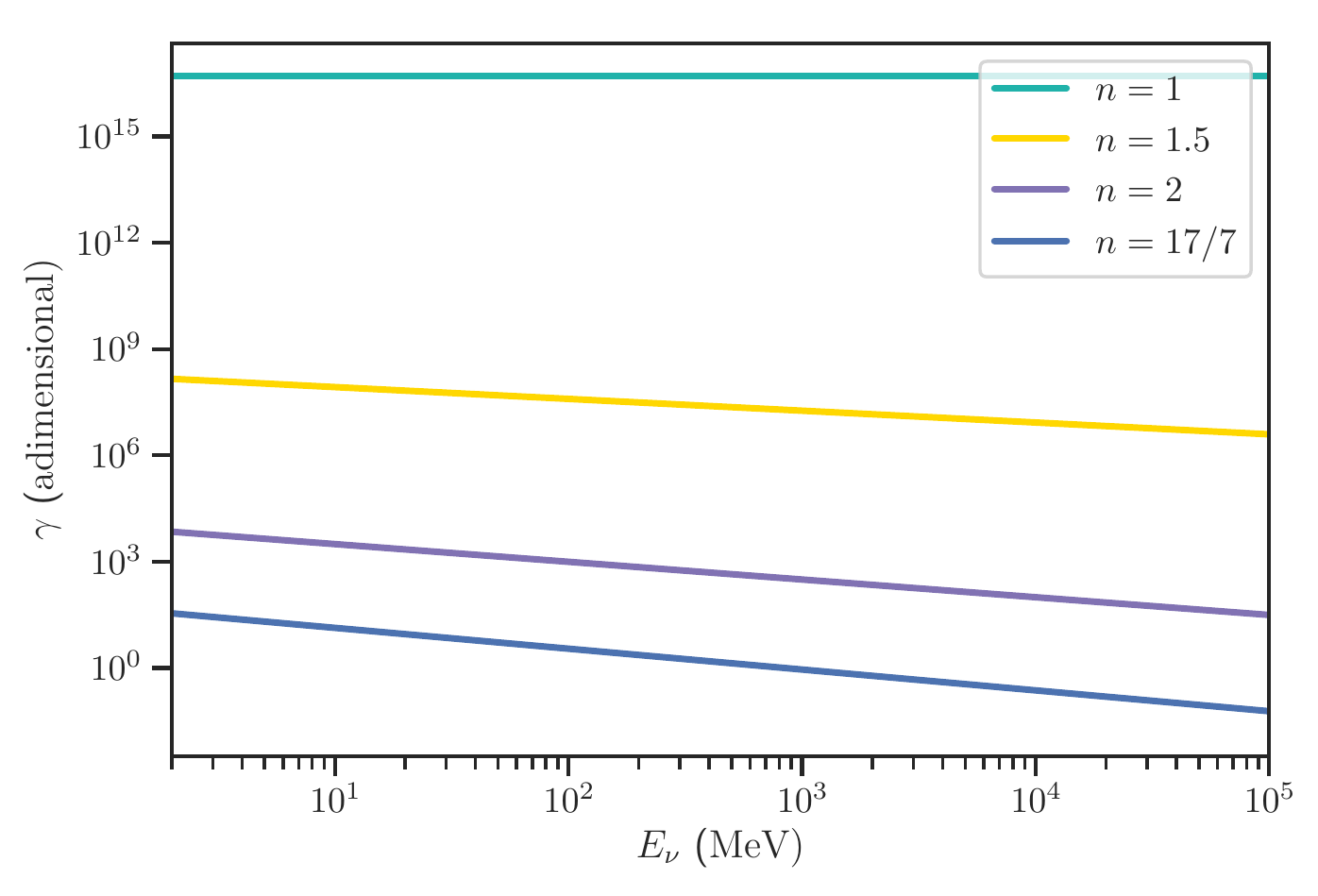}
    \caption{Adiabaticity parameter for a medium with density profile $\rho\propto  r^{-n}$ considering different exponents.}
    \label{fig:gamma_n}
\end{figure}

\begin{figure}
    \centering
    \includegraphics[width=0.45\textwidth]{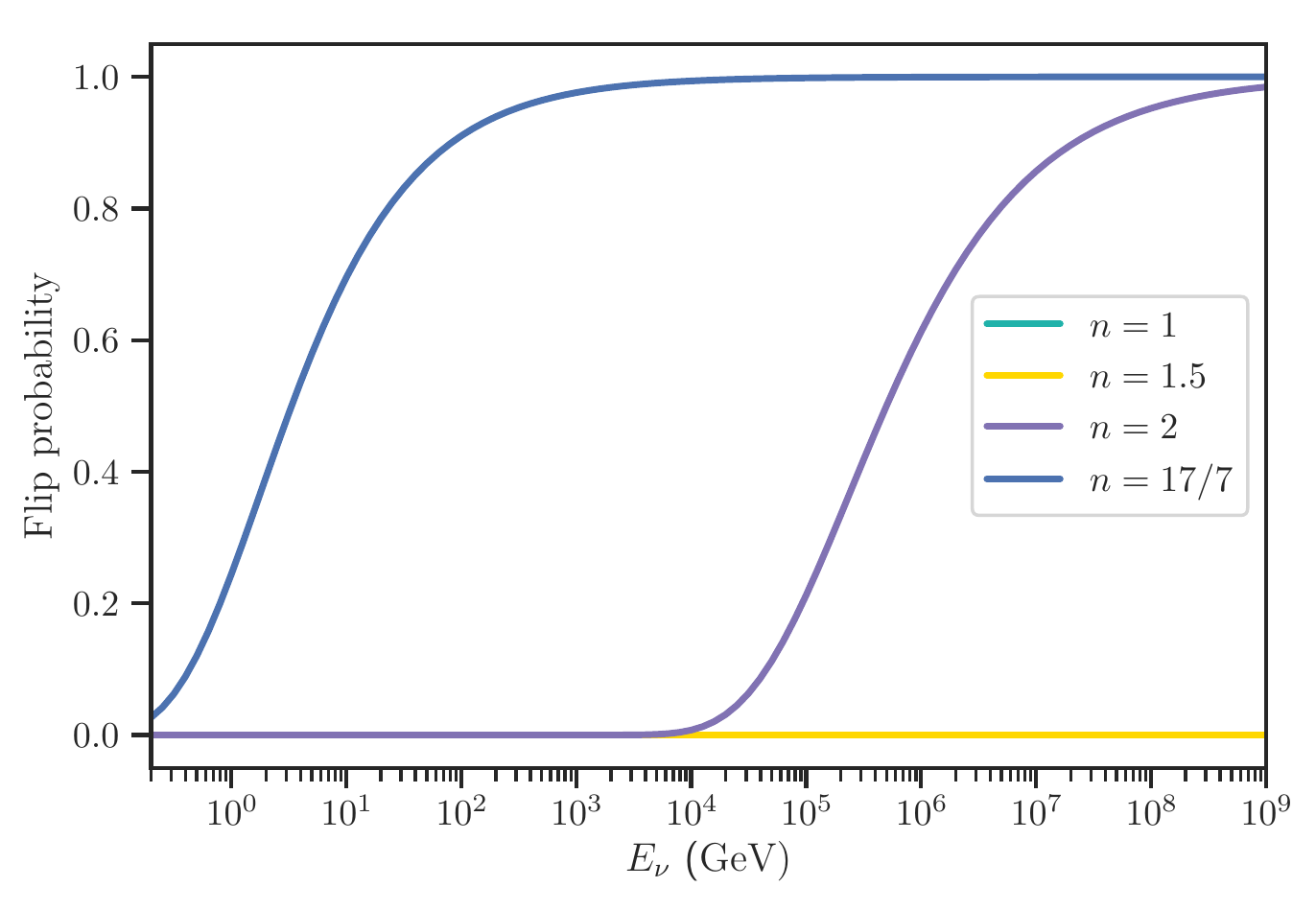}
    \caption{Flip probability obtained from the adiabaticity parameters in a previous Figure.}
    \label{fig:flip_n}
\end{figure}


Additionally, in Figure \ref{fig:gamma_n} we contemplate several power-law wind-profiles with typical exponents $(1,1.5,2,17/7)$ \citep{gra18, 2021ApJ...907...78F} to demonstrate that the adiabaticity condition is also satisfied. We find that slopes become steeper for density profiles with larger exponents, but they remain adiabatic in the MeV-range, and hence, their associated flip probabilities are also zero (see Figure \ref{fig:flip_n}).\\

\begin{figure*}
    \centering
    \includegraphics[width=0.97\textwidth]{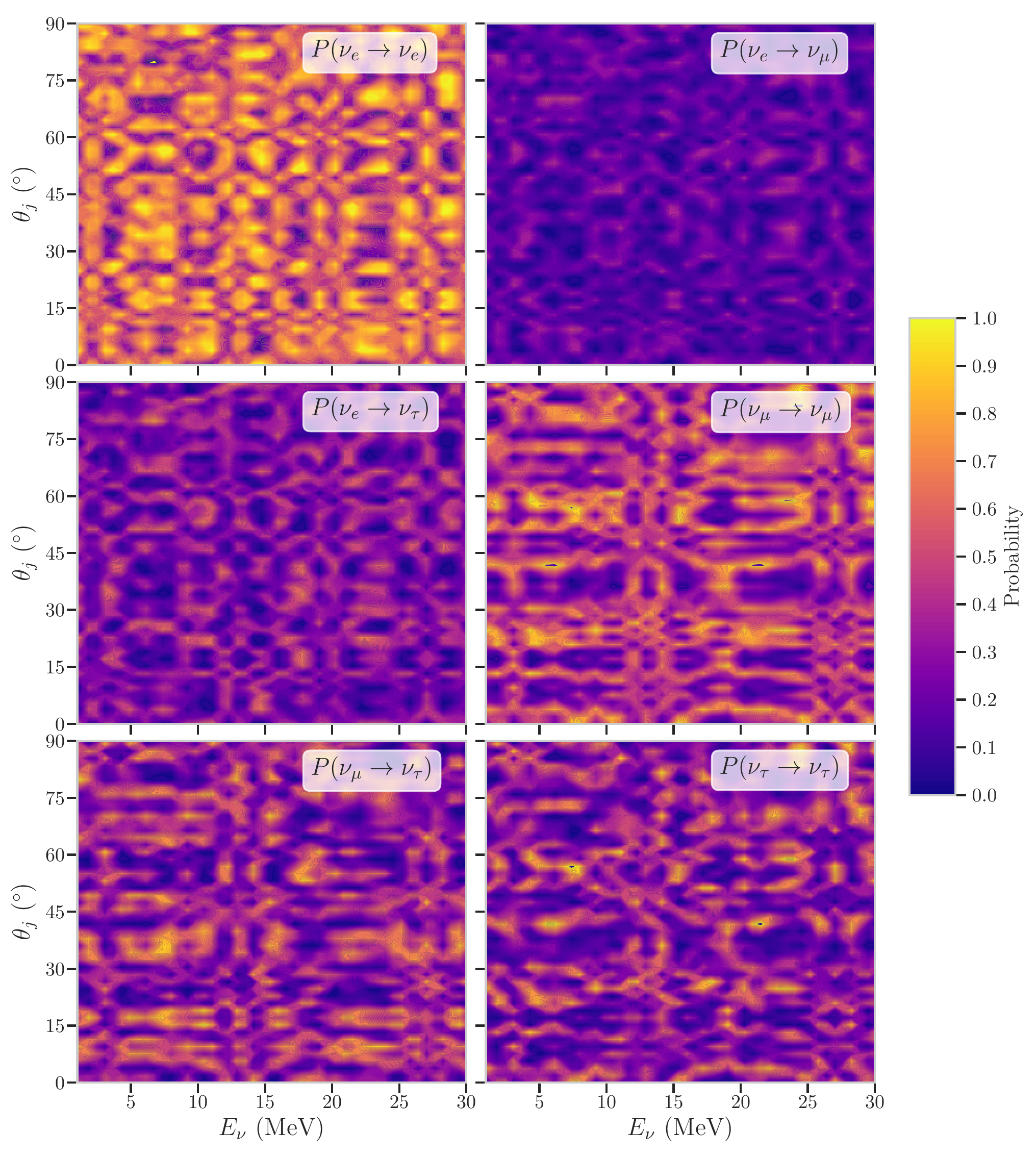}
    \caption{Oscillograms within a MDW scenario. We show the six independent possible transitions for this configuration. It is worth noting that the survival probability of the electronic neutrino dominates over the complementary probabilities.}
    \label{fig:magnetic_contourplot}
\end{figure*}

\begin{figure*}
    \centering
    \includegraphics[width=0.97\textwidth]{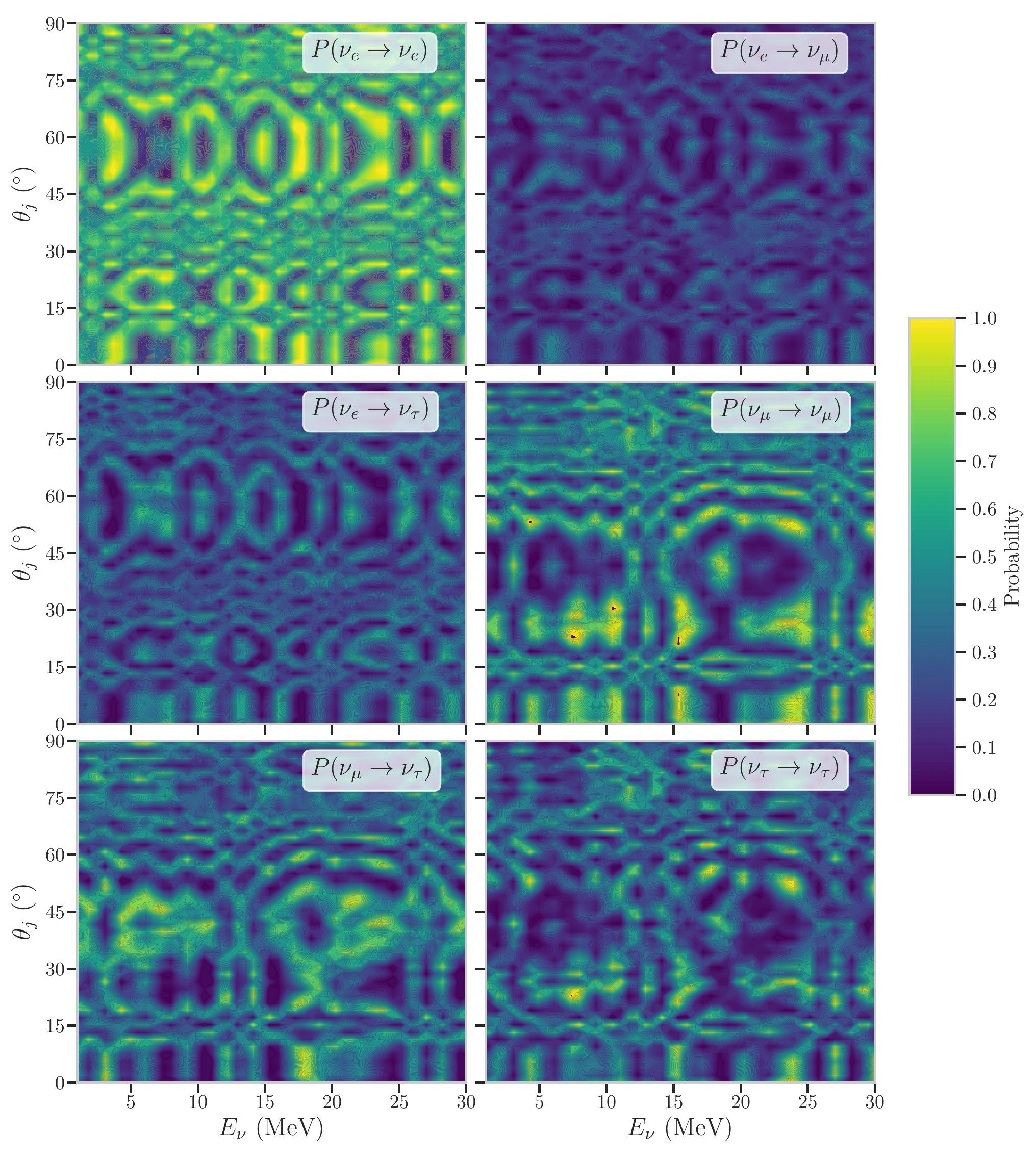}
    \caption{Same as in Figure \ref{fig:magnetic_contourplot}, but for a NDW scenario.}
    \label{fig:neutrino_contourplot}
\end{figure*}

\begin{figure*}
    \centering
    \includegraphics[width=0.97\textwidth]{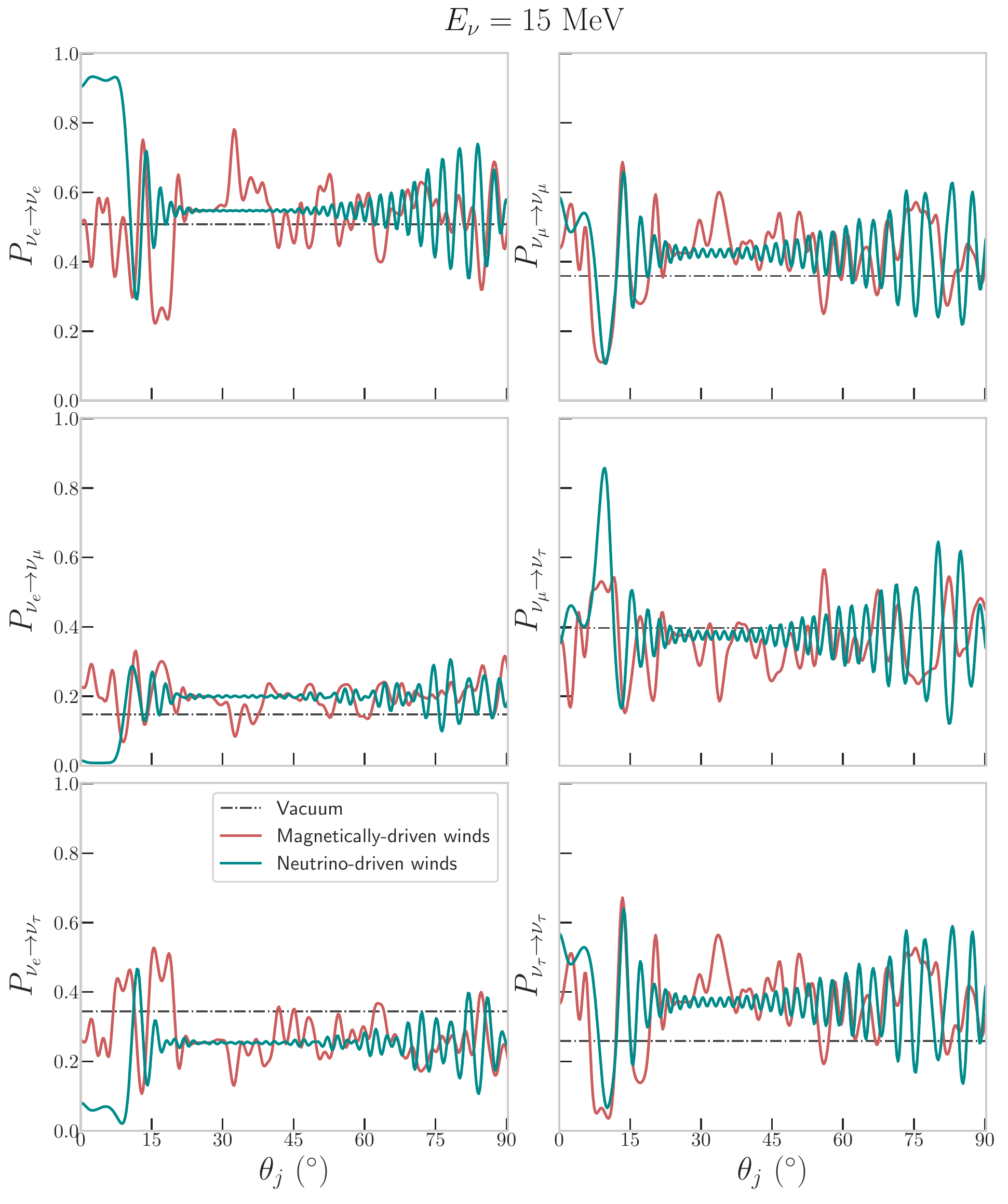}
    \caption{Oscillation probabilities projected on the $P-\theta_j$ plane for neutrinos propagating in the vacuum (gray dashed line), in a medium with winds dominated by magnetic processes (red line) and a medium with winds driven by neutrinos (blue line). One can distinguish the deviation from the standard oscillation probability in the vacuum for a 15$-$MeV neutrino. }
    \label{fig:proba_theta}
\end{figure*}

\begin{figure*}
    \centering
    \includegraphics[width=0.97\textwidth]{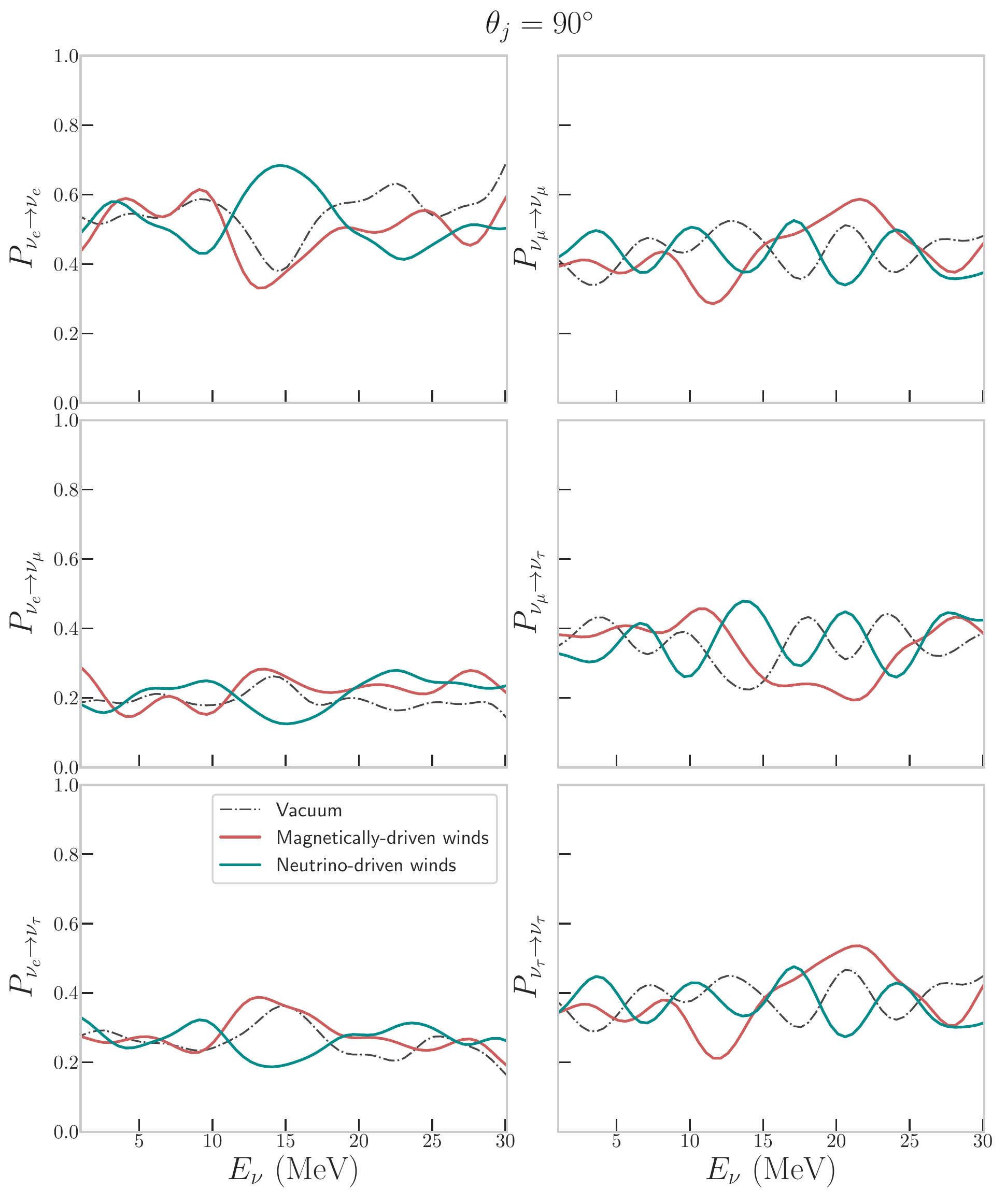}
    \caption{Oscillation probabilities projected on the $P-E_\nu$ plane for neutrinos propagating in a medium with winds dominated by magnetic processes (red line) and a medium with winds driven by neutrinos (blue line). For these plots we have to take into account a NO scheme. Here we consider MeV-neutrinos  in both scenarios propagating at the Equator ($\theta_j=90\ ^ \circ$) where the differences between the medium densities become significant. We also plot the oscillations in the vacuum for comparison purposes.}
    \label{fig:proba_energy}
\end{figure*}

In this way, from Equations (\ref{eq:proba_vacuum}) and (\ref{eq:Proba_m}), we calculate the oscillation probabilities at $r=10^9$ cm, taking into account an energy range between 1 and 30 MeV for both MDW and NDW scenarios. We also compare them with probabilities in the vacuum. To compute the exact matter oscillations probabilities numerically in a three-flavour admixture scenario, we use the general-purpose \texttt{NOPE} code \citep{bus19}, constructing the two Hamiltonians with their associated neutrino effective potential in each case. Our results are shown in  both Figures \ref{fig:magnetic_contourplot} and \ref{fig:neutrino_contourplot}, where we show the contour plots from each independent flavour transitions as a function of energy and $ \theta_j $. Since we want to study in more detail the changes in probabilities for neutrinos travelling in all directions,  we plot w.l.o.g.  in Figure \ref{fig:proba_theta}, the angular projection for neutrinos with $ E_\nu = 15 $ MeV. As a reference, we also compare them with neutrino oscillations in the vacuum, which, as expected, are constant for all $ \theta_j $ angles. We find slight variations in both cases concerning the vacuum with growing wiggles at lower and higher latitudes where oscillation amplitudes are larger because of boundary effects, being the mid-latitudes where the probabilities converge to an average value.  The same procedure was followed in Figure \ref{fig:proba_energy},  where we show the probabilities as a function of neutrino energy for a fixed angle of $\theta_j=90\ ^\circ$. Here, we notice a small shift between oscillation phases in the three contemplated cases. \\

\begin{figure*} 
\centering
	\subfloat
					{
  					\includegraphics[width=0.45\textwidth]{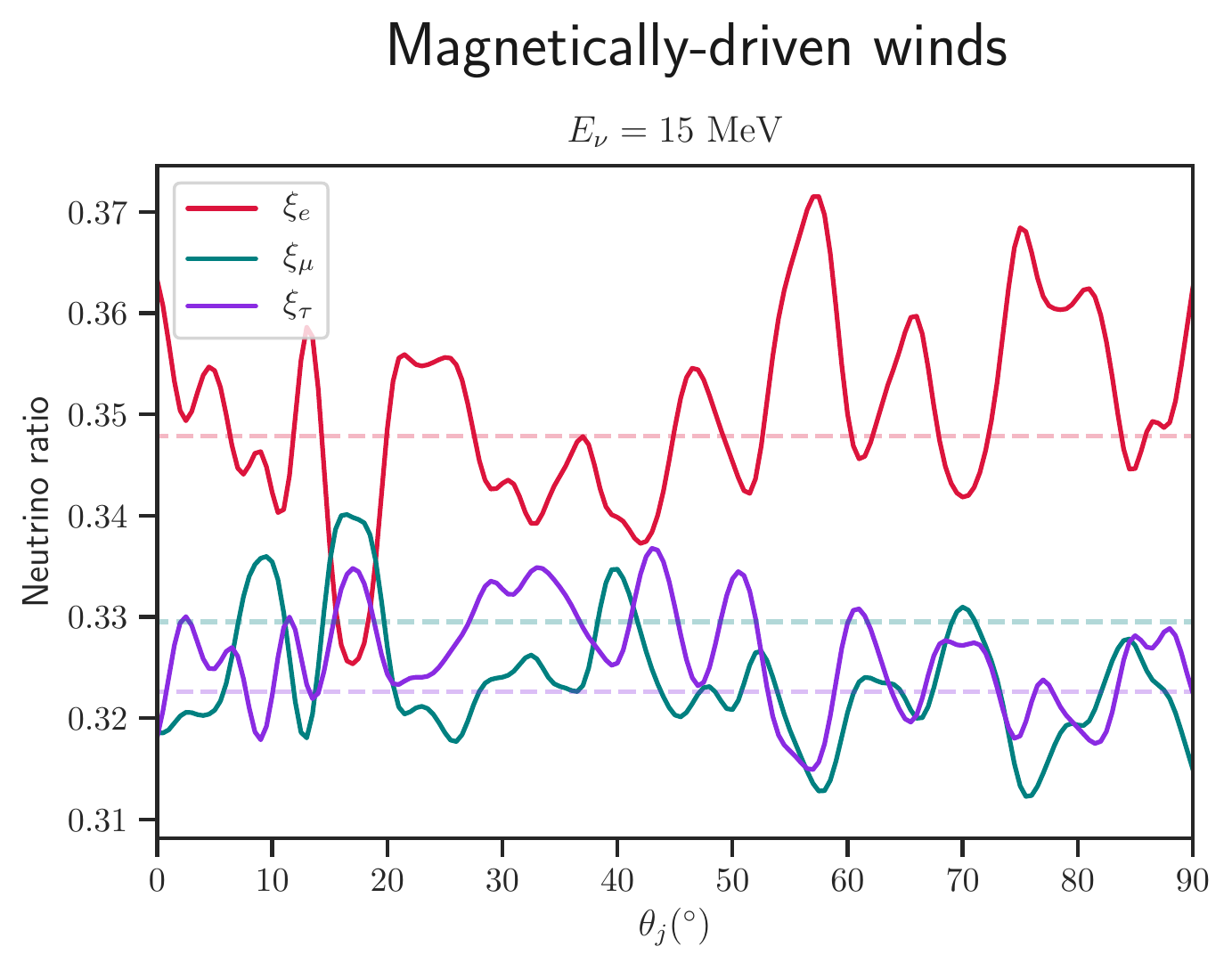}%
          	        }
					\qquad
	\subfloat
					{
  					\includegraphics[width=0.45\textwidth]{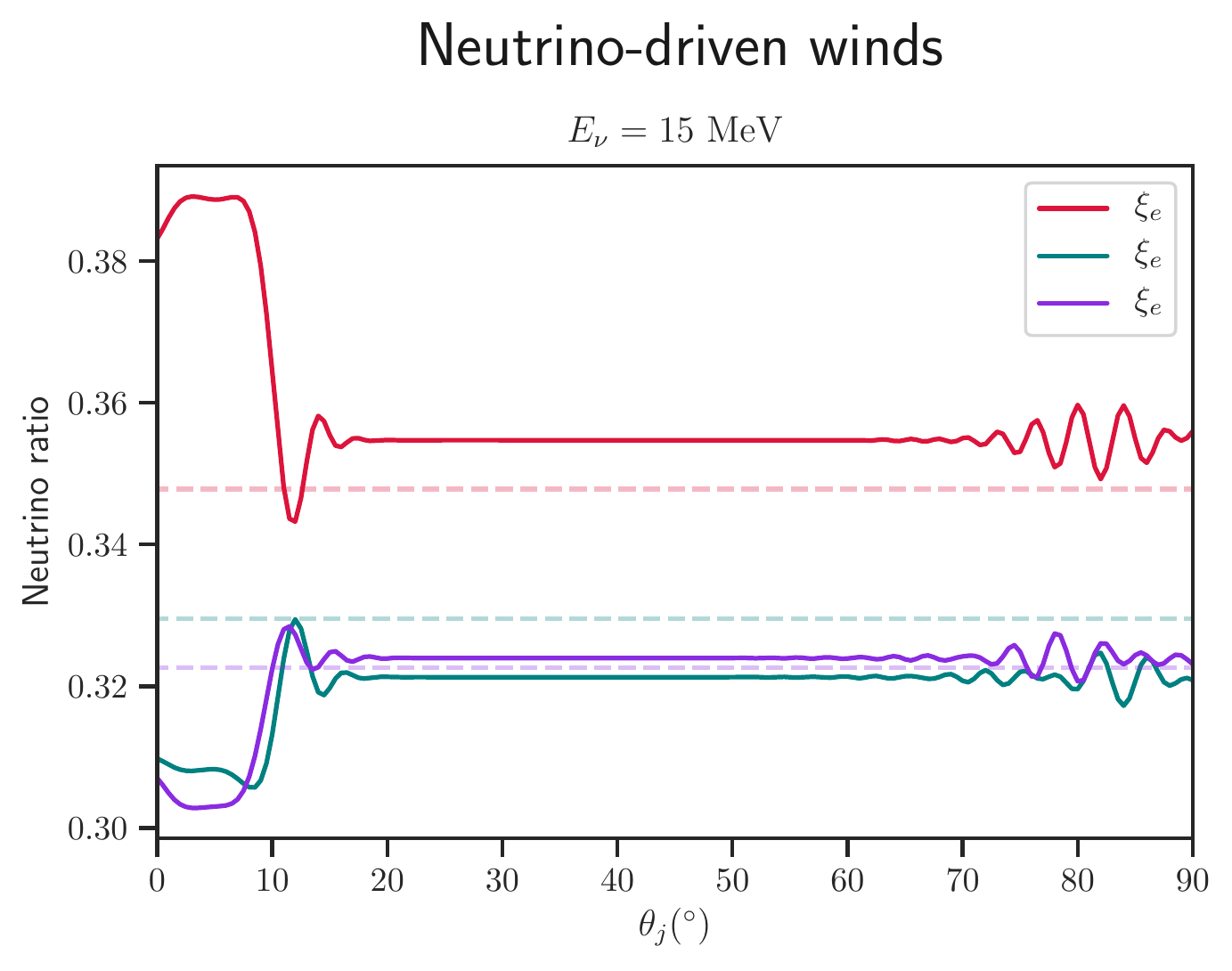}%
          	        } \qquad
	\subfloat
					{
  					\includegraphics[width=0.45\textwidth]{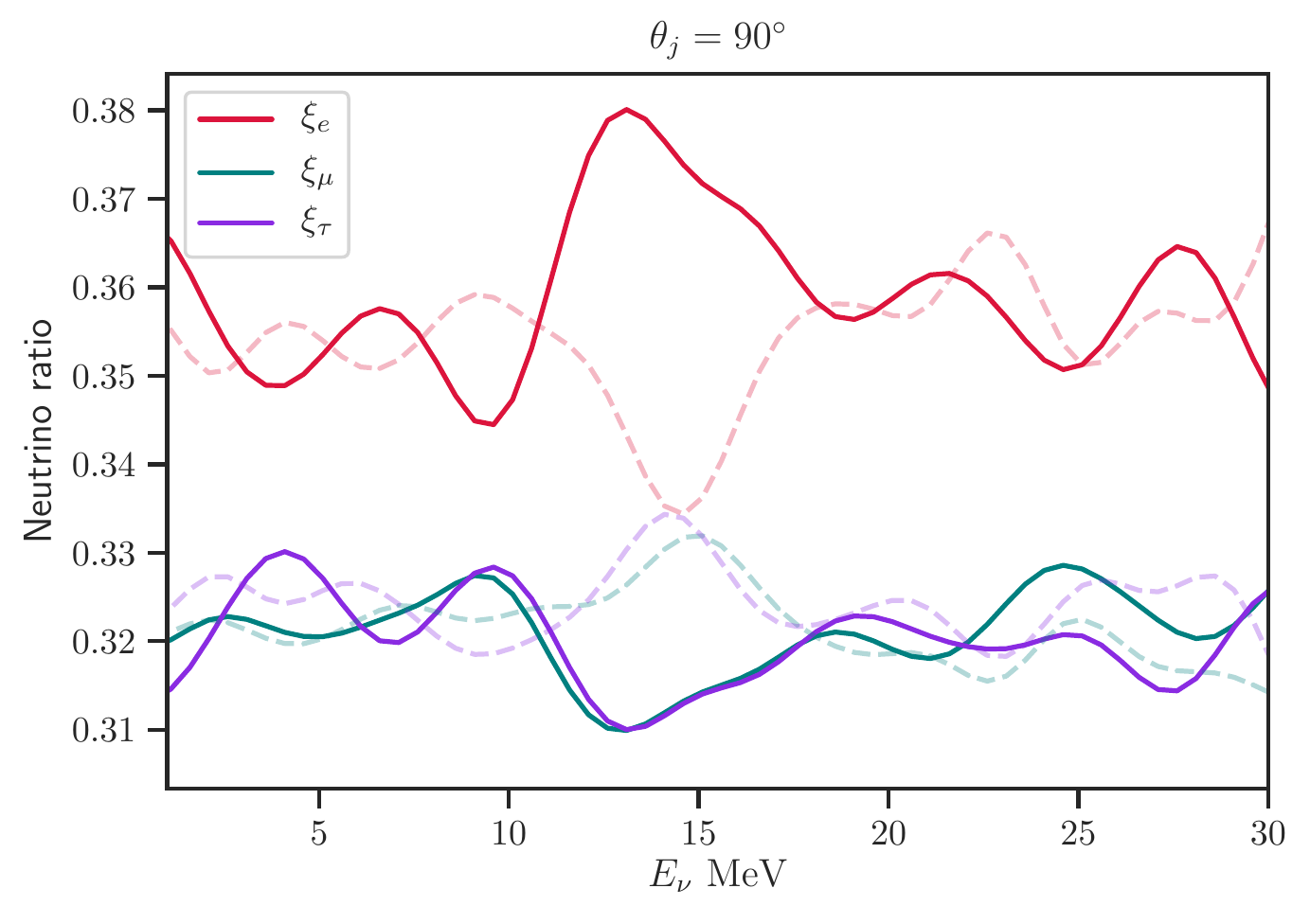}
         	        }
					\qquad
	\subfloat
					{
  					\includegraphics[width=0.45\textwidth]{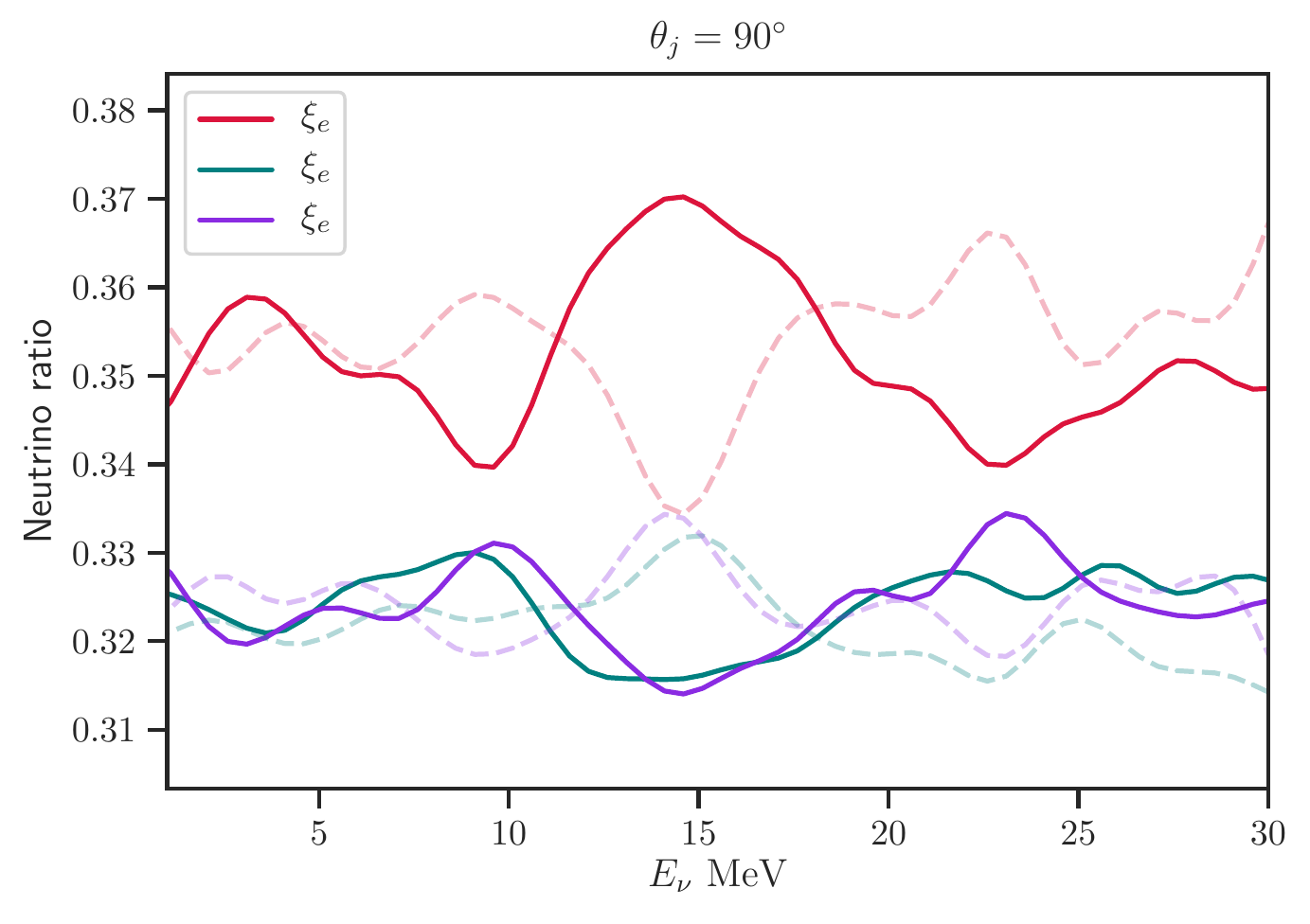}%
          	        }    
          	        \qquad  
          	        
	\subfloat
					{
  					\includegraphics[width=0.45\textwidth]{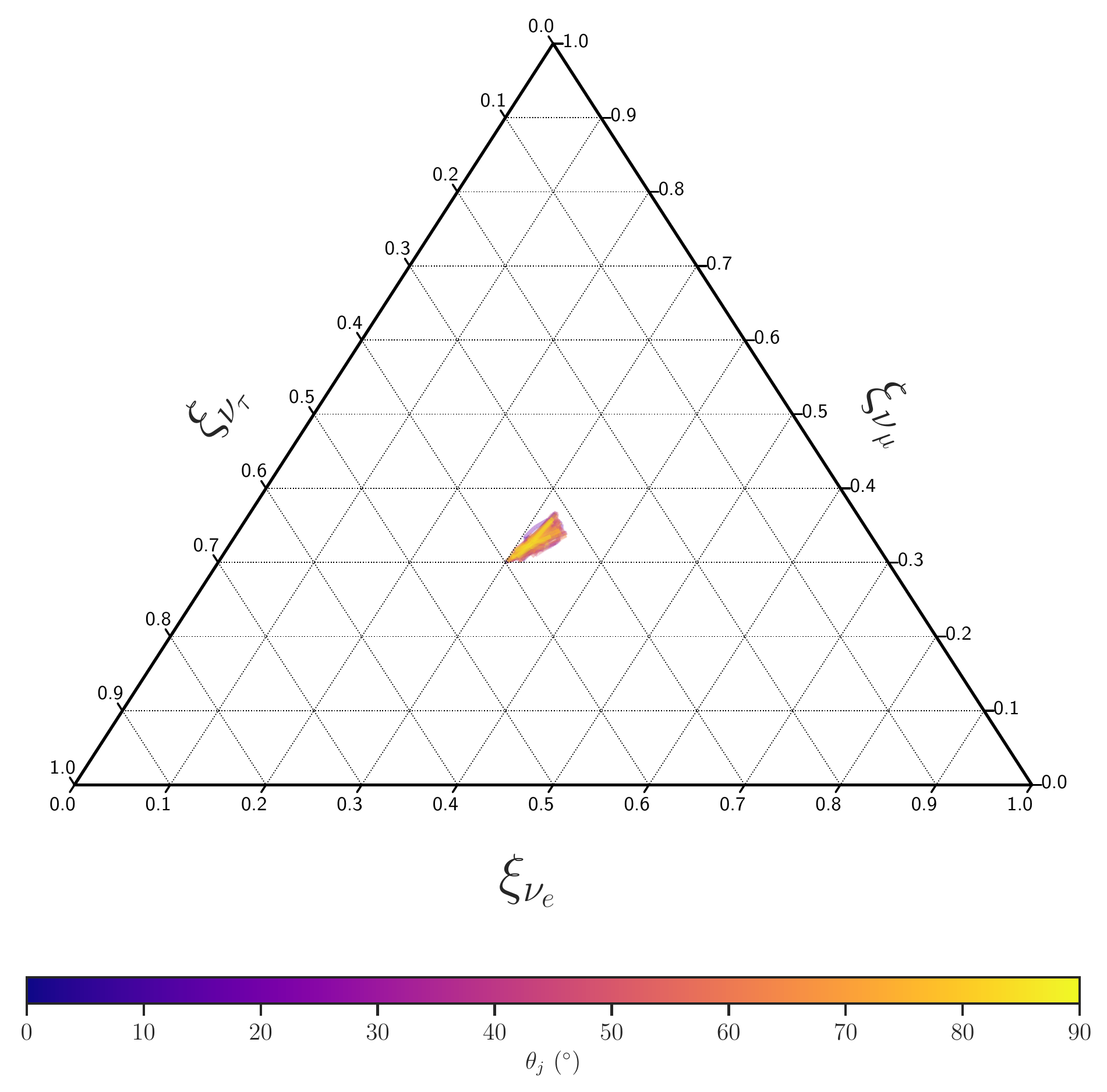}
          	        }
					\qquad
	\subfloat
					{
  					\includegraphics[width=0.45\textwidth]{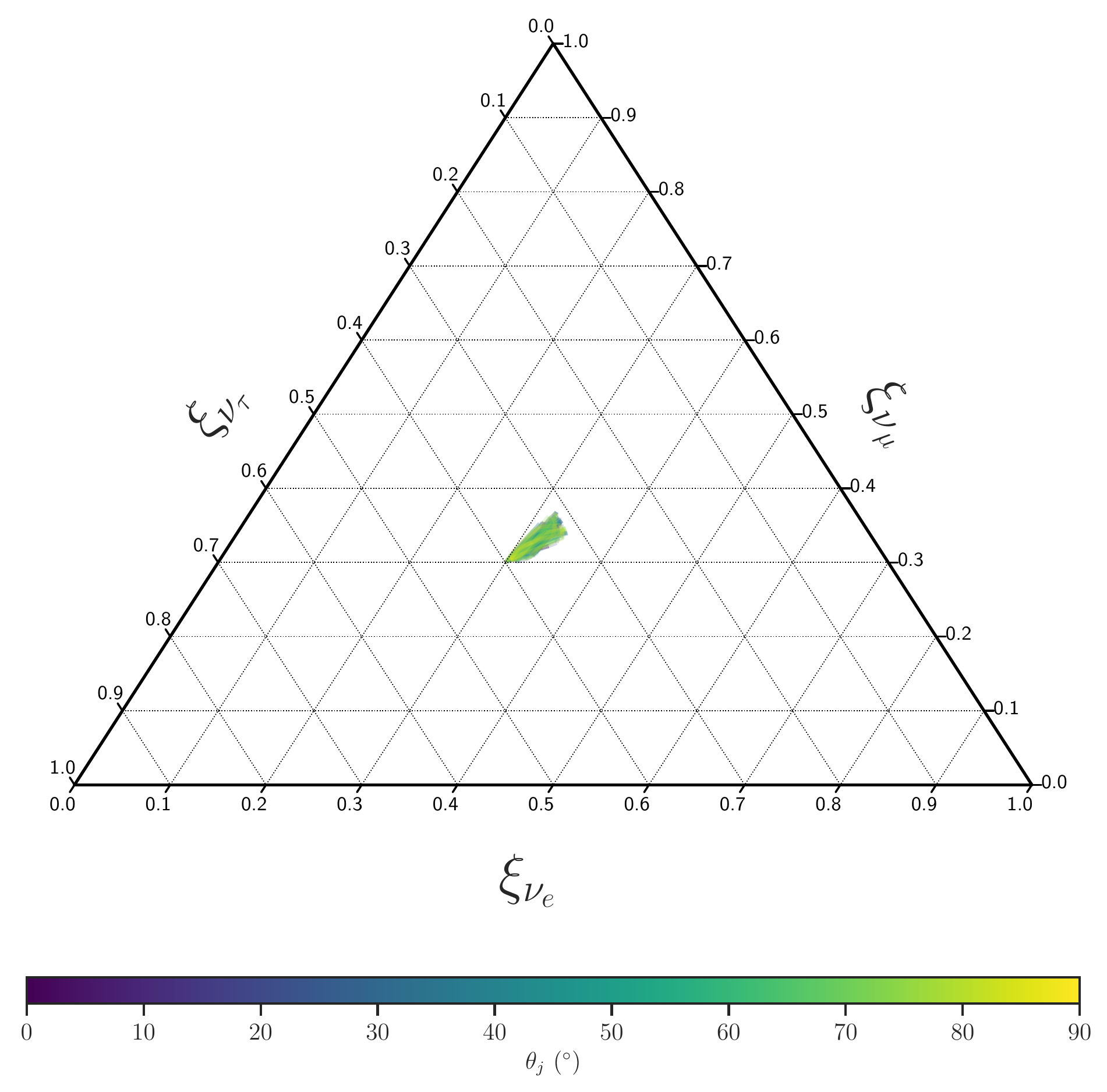}%
          	        }    
          	        \qquad            	        
\caption{\\\textbf{Left panels:} (Upper) Representation of the expected neutrino fractions (thick lines) within a MDW scenario as a function of $\theta_j$. Comparison with the theoretical neutrino ratios in the vacuum (dashed lines) is also shown. We notice a larger contribution of electronic flavour due to a overproduction of $\nu_e$ by charged-current interactions ($e^\pm$capture on nucleons) within the medium.\\ (Middle)  Same as above but as a function of $E_\nu$.\\ (Bottom) Ternary plot associated with this cases (bottom panel). We can observe that approximately the same neutrino ratio is conserved ($1/3:1/3:1/3$) and indeed, transitions only fluctuate between these allowed regions.
\\
\\
\textbf{Right panels:} Same as in left panels but for a NDW case. }  
\label{fig:ratio}        	        
\end{figure*}

Once we got the probabilities for each scheme, we proceed to calculate the neutrino rate using the parametrization shown in Equation (\ref{eq:xi}), assuming an initial creation rate of $(\xi_e,\xi_\mu,\xi_\tau)_{\rm created}\equiv(0.4,0.3,0.3)$. Therefore, we show in Figure \ref{fig:ratio}, the expected neutrino fraction as a function of energy and latitudinal angle and their associated ternary plot. In the left column, we represent the analysis performed for MDW winds, while in the right column, we perform the same but for NDW. \\

In both cases, we identify a defined oscillation region with boundaries between $\xi_{\rm min}=0.311$ and $\xi_{\rm max}=0.372$ in the $1-30$ MeV-range, suggesting that roughly the same flavour proportion $(1/3,1/3,1/3)$ is preserved.\\

Comparing the upper and middle panels, we found in both cases that angular as well as energy dependence is essential, and they play a crucial role in the oscillation performance. This effect is more prominent in the MDW case than in the NDW   because of the magnetic field amplification. Furthermore, we see that in the MDW case, the neutrino ratios do not converge to an expected value, while in the NDW case, the ratios only slightly deviate from the vacuum value. We could somehow use this condition to determine if the involved progenitor was a combination of NS-NS or BH-NS.

\section{Conclusions}\label{sec:conclusion}

Neutrinos are a valuable tool to characterize the astrophysical sources that produce them, mainly because they are created in a magnetized, very hot, and initially opaque medium \citep[e.g., see][]{2014MNRAS.437.2187F,2015MNRAS.450.2784F}. They propagate through a relatively large column density towards the Earth. In this work, we show that neutrino oscillation probabilities are affected when neutrinos cross a non-vacuum medium.  In particular, we found that admixtures and flavours are strongly dependent on the neutrino energies and density profiles. We derive for the first time the effective potential for neutrinos crossing this medium and compute the resulting oscillation properties when this effect is taken into account.\\

We found that dealing with a  $E_\nu=1.022$ MeV in the MDW case, the minimal angle from which oscillations become important is $\theta_j = 36.04\ ^\circ$, while the same neutrino but traveling within an NDW medium behaves similarly to the vacuum case through all directions, being $E_\nu=34.58 $ MeV the margin energy from which this effect is suppressed. Conversely, for more energetic neutrinos, such as, $E_\nu = 10$ MeV, the matter effects become relevant to angles as narrow as $\theta_j = 18.87\ ^\circ$.\\

Since the wind density depends mainly on the participating progenitors during the merger, we only considered MDW and NDW as the only two possible media types. In the first one, we assume that the medium has a remarkable magnetic contribution, which increases the baryonic density for higher latitudes and, as a consequence, the following characteristics stand out: i) at 90 degrees, the highest values are reached with $\rho=6.678\times10^{5}\ \rm{g cm}^{-3}$ and  $V_{\rm eff}=2.125\times10^{-9}$ eV, ii) resonance energy arises at $E_\nu = 14.07$ MeV, iii) the neutrino oscillation probabilities slightly deviate from the theoretical probability in the vacuum. For instance, the survival probability for a $15-$MeV electronic neutrino, propagating at $\theta = 35\ ^\circ$ in this medium is $0.382079$ while its counterpart in the vacuum is only $0.507873$, iv) the neutrino ratio in this medium for the same $15-$MeV neutrino is $\xi_{\rm MDW}=(0.335,0.337,0.328)$ in comparison with the vacuum value of $\xi_{\rm Vacuum}=(0.348,0.330,0.322)$.\\

In the second case, under a NDW approach, the following features arose: i)  $\rho=4.23\times10^{2}\ \rm{g cm}^{-3}$ and $V_{\rm eff}=1\times10^{-11}$ eV at equatorial latitudes, ii) we found a resonance energy of 28.97 MeV, iii) the same $15-$MeV electronic neutrino we used before has a surviving probability of $0.156094$ with a ratio of $\xi_{\rm NDW}=(0.314,0.346,0.340)$.\\

In both cases, the neutrino oscillations are within the adiabatic regime, between two well-defined regions dependent on the chosen initial production ratio. It is worth mentioning that we replicate the calculations considering both NO and IO mass hierarchies, but the differences were negligible in our energy range.\\

{Moreover, the neutrino ratio varies for a greater extent in the magnetic case. This effect is more predominant for higher latitude angles, while within the NDW regime, the expected ratio is only slightly altered from the vacuum case.} \\


{In conclusion, we can summarise that the winds produced during the formation of sGRB must be taken into account in the oscillation calculations since the resonant effects of thermal neutrinos within these media are essential. Even though it is possible to estimate the number of events that would be expected with future detectors within this energy range (e.g. the future Hyper-Kamiokande experiment), it would require studying the change in oscillation properties between the source, the intergalactic medium and the Earth itself, which is beyond this paper's scope. Nevertheless, it is possible to predict that if the conditions in which a GRB arises are optimal (sufficiently close and energetic with a promising flux of detectable neutrinos), in the next decades we will be able to detect neutrinos coming from this type of sources and cross-check the results presented in this work.}

\section*{Acknowledgements}
We thank J. Beacom, D. Page, and C. G. Bernal for useful discussions. GM acknowledges the financial support through the CONACyT grant 825482. NF acknowledges the financial support from UNAM-DGAPA-PAPIIT  through IN106521.

\section*{Data Availability}
The data underlying this article are primarily covered within the article. Further specific data will be shared on reasonable request to the corresponding authors.

\bibliographystyle{mnras}
\bibliography{Bib_osc} 


\clearpage

\end{document}